\documentclass[prb,twocolumn,floatfix,showpacs,preprintnumbers,superscriptaddress]{revtex4}
\usepackage{amsmath}
\usepackage{amssymb}
\usepackage{graphicx}
\usepackage{bm}
\usepackage{dcolumn}

\newcommand{\vso}{${\bm S}_0$}          
\newcommand{\mvso}{{\bm S}_0}           
\newcommand{\vs}{$\bm{S}$}              
\newcommand{\mvs}{\bm{S}}               
\newcommand{\mhat}{$\hat{\bm{m}}$}      
\newcommand{\mmhat}{\hat{\bm{m}}}       
\newcommand{\bhat}[1]{\hat{\bm{#1}}}    
\newcommand{\so}{$S_0$}                 
\newcommand{\etaso}{$\eta S_0$}         
\newcommand{\vbapp}{${\bm B}_{app}$}    
\newcommand{\mvbapp}{{\bm B}_{app}}     
\newcommand{\bapp}{$B_{app}$}           
\newcommand{\mbapp}{B_{app}}            
\newcommand{\Bn}{$B_N$}                 
\newcommand{\mvBn}{{\bm B}_N}           
\newcommand{\bn}{$b_N$}                 
\newcommand{\mva}{{\bm A}}              
\newcommand{\bhalf}{$B_{1/2}$}          
\newcommand{\mbhalf}{B_{1/2}}           
\newcommand{\xibl}{$\sqrt{\xi}B_L$}     
\newcommand{\mxibl}{\sqrt{\xi}B_L}      
\newcommand{\mxiblsq}{\xi B_L^2}        
\newcommand{\xiblsq}{$\mxiblsq$}        
\newcommand{\tpol}{$T_{pol}$}           
\newcommand{\tone}{$T_1$}               
\newcommand{\mtonerate}{T_1^{-1}}       
\newcommand{\tonerate}{$\mtonerate$}    
\newcommand{\degrees}{$^\circ$}         
\newcommand{\etal}{~\textit{et~al.}}   
\newcommand{\vbone}{${\bm B}_1$}        
\newcommand{\bone}{$B_1$}               
\newcommand{\bprime}{$B_{app}^\prime$}  
\newcommand{\vbprime}{$\bm{B}_{app}^\prime$}    
\newcommand{\mbprime}{B_{app}^\prime}   
\newcommand{\AlGaAs}{$\text{Al}_{0.1}\text{Ga}_{0.9}\text{As}$}
\newcommand{\AlGaAsX}{$\text{Al}_{x}\text{Ga}_{1-x}\text{As}$}
\newcommand{\mfl}{f_\ell}               
\newcommand{\fl}{$\mfl$}                
\newcommand{\mIinit}{{\bm I}_{av}^\text{\,initial}} 
\newcommand{\Iinit}{$\mIinit$}          
\newcommand{\mIprime}{{\bm I}_{av}^\prime}  
\newcommand{\Iprime}{$\mIprime$}        
\newcommand{\gfctr}{$\mathrm{g}$-factor}    
\newcommand{\mgfctrEff}{\mathrm{g}^*}   

\begin{document}


\title{Electron Spin Dynamics and Hyperfine Interactions in Fe/{\AlGaAs}/GaAs Spin Injection Heterostructures}

\author{J.~Strand}
\affiliation{School of Physics and Astronomy}
\author{X.~Lou}
\affiliation{School of Physics and Astronomy}
\author{C.~Adelmann}
\affiliation{Department of Chemical Engineering and Materials
Science, \\University of Minnesota, Minneapolis, MN 55455}
\author{B.~D.~Schultz}
\affiliation{Department of Chemical Engineering and Materials
Science, \\University of Minnesota, Minneapolis, MN 55455}
\author{A.~F.~Isakovic}
\altaffiliation[Current address: ]{Laboratory of Atomic and Solid-State Physics, Cornell University, Ithaca, NY 14853} 
\affiliation{School of Physics and Astronomy}
\author{C.~J.~Palmstr{\o}m}
\affiliation{Department of Chemical Engineering and Materials
Science, \\University of Minnesota, Minneapolis, MN 55455}
\author{P.~A.~Crowell}\email{crowell@physics.umn.edu}
\affiliation{School of Physics and Astronomy}

\date{\today}

\begin{abstract}
We have studied hyperfine interactions between spin-polarized electrons and lattice nuclei in \AlGaAs/GaAs quantum well (QW) heterostructures.  The spin-polarized electrons are electrically injected into the semiconductor heterostructure from a metallic ferromagnet across a Schottky tunnel barrier.  The spin-polarized electron current dynamically polarizes the nuclei in the QW, and the polarized nuclei in turn alter the electron spin dynamics.  The steady-state electron spin is detected via the circular polarization of the emitted electroluminescence.  The nuclear polarization and electron spin dynamics are accurately modeled using the formalism of optical orientation in GaAs.  The nuclear spin polarization in the QW is found to depend strongly on the electron spin polarization in the QW, but only weakly on the electron density in the QW.  We are able to observe nuclear magnetic resonance (NMR) at low applied magnetic fields on the order of a few hundred Oe by electrically modulating the spin injected into the QW.  The electrically driven NMR demonstrates explicitly the existence of a Knight field felt by the nuclei due to the electron spin.
\end{abstract}


\maketitle

\section{\label{Introduction} Introduction}

Spin-polarized electrons in GaAs interact with lattice nuclei through the hyperfine interaction, leading to dynamic nuclear polarization (DNP).\cite{Lampel:First-opt-DNP, OO, Paget:GaAsNuclearSpinCoupling-PRB1977}  Typical DNP experiments in GaAs exploit the spin-dependent selection rules for optical transitions to generate the necessary population of spin-polarized electrons in the conduction band.  Recent experiments have demonstrated the electrical injection of spin-polarized electrons using a metallic ferromagnet as a contact and a band-engineered Schottky tunnel barrier.\cite{Ploog:Zhu:first-Fe-spinLED-PRL2001, Jonker:Hanbicki:Fe-AlGaAsAPL2002, Jonker:VanTErve:AlO-Schottky2004, Jonker:Li:110-Fe-GaAs-SpinLED-APL2004, IMEC:Motsnyi:OHE-APL2002, IMEC:motsnyi:PRB2003, IMEC:VanDorpe:AlO-Schottky2004, Parkin:Jiang:MTT-spinLED-PRL2003, Crowell:Strand-spinLED-DNP-PRL2003, Crowell:Strand-spinLED-DNP-NMR-APL2003, Crowell:Adelmann-spinLED-BiasDep-PRL2004}  The question naturally arises whether the electrically injected population of spin-polarized carriers interacts with lattice nuclei in a manner similar to optically injected spin-polarized carriers.  In this paper we present a series of experiments in which a current of spin-polarized electrons electrically injected from an Fe contact dynamically polarizes lattice nuclei in a \AlGaAs/GaAs quantum well (QW). The nuclear polarization and its subsequent effects on electron spin dynamics are well-described by the same model of DNP used to understand optical orientation experiments.  However, the magnetic anisotropies of the Fe contact allow access to new geometries for observing the interaction between spin-polarized electrons and nuclei that cannot be achieved in optical spin-pumping experiments.  In addition, we show that modulation of the spin polarization of the current can be used to control the hyperfine interaction directly, leading  to the observation of nuclear magnetic resonance in very low applied fields.

\section{\label{sec:background}Spin Injection: Background}

Many experiments have utilized the selection rules for optical transitions in GaAs to measure the spin polarization of conduction band electrons.\cite{Fiederling:Molenkamp:FirstBeMnZnSe-spinLED, OhnoY:Awschalom:FirstGaMaAs_spinLED, OO, Awsch:SpintronicsBook, zutic:SpintronicsRMP}  These selection rules allow for a simple mapping between the net luminescence circular polarization $P_{EL}=(I_+-I_-)/(I_++I_-)$ where $I_{+,-}$ are the intensities of the two helicities of circularly polarized light, and the average electron spin along the sample normal: $P_{spin}=\alpha S_z=\alpha \mvs\cdot\hat{\bm{z}}$, where {\vs} is the average electron spin in the quantum well (QW), and $\alpha=2$ for luminescence from QW systems while $\alpha=1$ for bulk systems.\cite{OO:DyakonovPerel}$^,$\cite{en:Value-of-s} The orientation and magnitude of {\vs} will in turn depend on the applied magnetic field {\vbapp} and the electron spin lifetime $T_s=(\tau_s^{-1}+\tau^{-1})^{-1}$, where $\tau_s^{-1}$ is the rate of electron spin relaxation and $\tau^{-1}$ is the rate of electron-hole recombination.  The electron spin immediately after injection into the QW is given by the vector {\vso}.   Spin relaxation that occurs before the electrons recombine with holes reduces the total spin in the QW from its initial value {\so} to its steady-state value $S$.  In addition, the spin will precess about {\vbapp} whenever $\mvbapp\times\mvs\neq0$.  The angle through which the spin precesses is determined by the product $\bm{\Omega}T_s$ of the Larmor precession frequency and the electron spin lifetime where ${\bm \Omega}=\mathrm{g}^*\mu_B{\bm B}_{app}/\hbar$, $\mathrm{g}^*$ is the effective electron $\mathrm{g}$-factor, $\mu_B$ is the Bohr magneton and $\hbar$ is Planck's constant. Typically, $\Omega\sim10^{10}\text{~GHz}$ at 5~kG and $T_s\sim200$~ps, giving $\Omega T_s\lesssim2\pi$. Combining these processes of injection, precession, relaxation and recombination, we can define a rate equation that describes the dynamics of the electron spin in the QW:\cite{OO:DyakonovPerel}
\begin{equation}\label{eq:spin-rate-eqn}
\frac{d\mvs}{dt}=\bm{\Omega}\times\mvs-\frac{\mvs}{\tau_s}-\frac{\mvs-\mvs_0}{\tau}.
\end{equation}
Luminescence is a steady-state measurement, and therefore we can set $d\mvs/dt=0$ and solve for {\vs}.  It is convenient to define the characteristic magnetic field $B_{1/2}=\hbar/(\mathrm{g}^*\mu_B T_s)$ at which $\Omega_{B_{1/2}}T_s=1$.  The steady-state solution of Eq.~\ref{eq:spin-rate-eqn} can then be written as
\begin{equation}
   {\bm S}=\eta \frac{B_{1/2}^2{\bm S}_0+({\bm S}_0 \cdot \mvbapp)\mvbapp+B_{1/2}(\mvbapp\times{\bm S}_0)}{B_{1/2}^2+B_{app}^2},
\label{eq:S-vector-no-Nuc}
\end{equation}
where $\eta=1/(1+\tau/\tau_s)$.\cite{OO:FleischerMerkulov}
Equation~\ref{eq:S-vector-no-Nuc} is the general equation for electron spin dynamics in GaAs under steady-state conditions. For the experiments under discussion here, the $z$-component $S_z$ is detected by the luminescence circular polarization $P_{EL}$, and {\vso} is electrically injected from a ferromagnetic metal contact.

In the case of electrical injection of spin-polarized electrons from a thin Fe film, $\mvso=\epsilon\rho\mmhat$, where {\mhat} is the direction of the Fe contact magnetization, $\rho$ is the spin polarization of the Fe contact, and $\epsilon$ is the efficiency of spin transport across the interface. Magnetic shape anisotropy in the thin film causes the Fe magnetization to lie in-plane at low magnetic fields.  Hence, at $\mvbapp=0$, {\vso} will be entirely in-plane and  the luminescence along the $z$-direction will be unpolarized. In order to achieve $S_z\neq 0$ in the QW, either the Fe magnetization must be rotated out of the plane prior to the injection of spin-polarized electrons, or the injected spin must precess out-of-plane after reaching the QW.  The first approach can be accomplished by applying a magnetic field along the $z$-direction in the longitudinal, or \textit{Faraday} geometry,\cite{Fiederling:Molenkamp:FirstBeMnZnSe-spinLED, Jonker:ZnMnSe-AlGaAsSpinLED-PRB2000, Ploog:Zhu:first-Fe-spinLED-PRL2001, Jonker:Hanbicki:Fe-AlGaAsAPL2002, Jonker:VanTErve:AlO-Schottky2004, Jonker:Li:110-Fe-GaAs-SpinLED-APL2004, Parkin:Jiang:MTT-spinLED-PRL2003, Crowell:Adelmann-spinLED-BiasDep-PRL2004} while the second requires an angle between {\vso} and {\vbapp} in the transverse, or \textit{Voigt} geometry.\cite{IMEC:Motsnyi:OHE-APL2002, IMEC:motsnyi:PRB2003, IMEC:VanDorpe:AlO-Schottky2004, Crowell:Strand-spinLED-DNP-PRL2003, Crowell:Strand-spinLED-DNP-NMR-APL2003}

\section{\label{sec:SamplePrep}Sample design, growth and processing}

\begin{figure}\begin{center}
    \includegraphics*{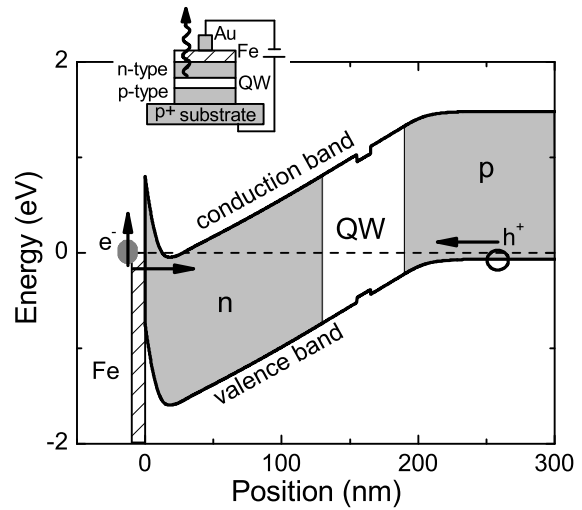}
    \caption[Spin-LED band structure and device schematic.]{\label{fig:spinLED-basics}Spin-LED band structure at zero bias\cite{SimWindows} and a schematic cross section  (inset) of the spin-LED device. The dashed line is the Fermi level, the open circle and arrow represents the flow of unpolarized holes from the p-type layer to the QW, the filled circle represents the flow of spin-polarized electrons injected from the Fe contact through the Schottky barrier.}
\end{center}\end{figure}

The devices studied here combine electrical injection of spin-polarized carriers across an engineered Schottky barrier with optical spin detection.  This device design is referred to as a spin-sensitive light emitting diode (LED), or spin-LED.\cite{Fiederling:Molenkamp:FirstBeMnZnSe-spinLED, OhnoY:Awschalom:FirstGaMaAs_spinLED, Jonker:ZnMnSe-AlGaAsSpinLED-PRB2000, Ploog:Zhu:first-Fe-spinLED-PRL2001, Jonker:Hanbicki:Fe-AlGaAsAPL2002, Jonker:VanTErve:AlO-Schottky2004, Jonker:Li:110-Fe-GaAs-SpinLED-APL2004, IMEC:Motsnyi:OHE-APL2002, IMEC:motsnyi:PRB2003, IMEC:VanDorpe:AlO-Schottky2004, Parkin:Jiang:MTT-spinLED-PRL2003, Crowell:Strand-spinLED-DNP-PRL2003, Crowell:Strand-spinLED-DNP-NMR-APL2003, Crowell:Adelmann-spinLED-BiasDep-PRL2004}  A vertical block diagram schematic of the spin-LED is shown in the inset of Fig.~\ref{fig:spinLED-basics}.  
A ferromagnetic metal is deposited on top of a highly doped n+ layer of {\AlGaAs}. Underneath the n+ layer is a drift layer of n-type {\AlGaAs} that separates the injector region of the device from the detector region of the device.  The detector is a 100~{\AA} GaAs QW with un-doped {\AlGaAs} barriers, positioned approximately 1500~{\AA} from the ferromagnet/semiconductor interface.  Beneath the QW detector is a layer of p-doped {\AlGaAs} that serves as a source of unpolarized holes with which the injected electrons recombine in the QW.  The device is operated with the bias applied between the ferromagnet and the substrate, and luminescence is collected along the growth direction $\bhat{z}$ of the sample.

\begingroup
\begin{table*}
\begin{ruledtabular} 
\begin{tabular}{crl|cl}
\multicolumn{3}{c}{Sample A}       & \multicolumn{2}{c}{Sample B}      \\
\multicolumn{2}{l}{thickness (nm)}&\multicolumn{1}{l}{material}&\multicolumn{2}{l}{thickness (nm)}\\ \hline
2.5  & $\text{T}_\text{growth}\sim 0$~$^\circ$C &Al& \multicolumn{2}{c}{\textit{same}}\\
5    & $\text{T}_\text{growth}\sim 0$~$^\circ$C &Fe&  \multicolumn{2}{c}{\textit{same}}\\
15   & n+ ($5\times 10^{18}\text{ cm}^{-3}$)&    {\AlGaAs}       &2.5 & i \\
15   &n/n+ (graded doping)   & {\AlGaAs}  &     \multicolumn{1}{r}{$\delta$-doped (Si)}&  n+ ($3\times 10^{13}\text{ cm}^{-2}$)\\
100  &n ($1\times 10^{16}\text{ cm}^{-3}$)& {\AlGaAs} &      100 & n ($6.7\times 10^{16}\text{ cm}^{-3}$)\\
25   &i  & {\AlGaAs}  &      \multicolumn{2}{c}{\textit{same}}\\
10   &i-QW &  GaAs    &     \multicolumn{2}{c}{\textit{same}}\\
25   &i   &{\AlGaAs}  &      \multicolumn{2}{c}{\textit{same}}\\
50   &p/p+ ($1\times10^{17}-1\times 10^{18}\text{ cm}^{-3}$)&   {\AlGaAs} & \multicolumn{2}{c}{\textit{same}}\\
150  &p+ ($1\times 10^{18}\text{ cm}^{-3}$)&    {\AlGaAs}  &  \multicolumn{2}{c}{\textit{same}}\\
300  &p+ ($1.1\times 10^{18}\text{ cm}^{-3}$)& GaAs     & \multicolumn{2}{c}{\textit{same}}\\
\multicolumn{1}{r}{substrate}   &p+ &    GaAs (100)   &     \multicolumn{2}{c}{\textit{same}}
\end{tabular}
\end{ruledtabular}
\caption[Spin-LED heterostructure details]{\label{tbl:sample-differences}Heterostructure details for the two primary spin-LED samples discussed in this paper.}
\end{table*}
\endgroup

 The spin-LED heterostructure consists of two back-to-back diodes: a Schottky diode at the interface (injector) followed by
a n-i-p light emitting diode (detector).  The device is operated with the Schottky contact reverse biased (electrons passing from Fe into {\AlGaAs}) and the n-i-p LED forward biased.  Positive device voltages in this paper refer to reverse Schottky and forward LED bias conditions: electrons tunnelling into the Al$_{0.1}$Ga$_{0.9}$As and recombining with holes in the QW to emit photons.  Transport at the interface is determined by the Schottky barrier shape and width, which in turn depends on the {\AlGaAs} doping at the interface. The combination of the ferromagnetic film and highly doped n+ interfacial Al$_{0.1}$Ga$_{0.9}$As is referred to as the
injector, while the n-i(QW)i-p LED is referred to as the detector. 

 Two different injector doping designs are used for the data presented here: graded doping and $\delta$-doping.  Sample A has a graded doping injector similar to that of Hanbicki{\etal}\cite{Jonker:Hanbicki:Fe-AlGaAsAPL2002} and is used for all field dependence and time dependence measurements in this paper.  Sample B has a $\delta$-doped injector and is used for nuclear magnetic resonance measurements.  The $\delta$-doped injector is created by depositing a two-dimensional sheet of n-type dopant (Si) near the interface.   Table~\ref{tbl:sample-differences} shows details of these two heterostructure designs.  Several samples with
varying $\delta$-doping, graded interfacial doping, drift layer doping, and QW doping have been grown and tested and exhibit varying degrees electrical spin injection and dynamic nuclear polarization.

The samples are grown using molecular beam epitaxy (MBE).  The Fe film is deposited \textit{in situ} and is capped with a thin Al layer to prevent oxidation.  The devices are processed using standard photolithography and wet etching.  Mesas, either round or rectangular, are lithographically defined and etched below the level of the n-i-p depletion region to minimize leakage currents.  The samples are mounted in chip carriers using indium solder (diffused into the backside of the wafer prior to growth) that creates an ohmic contact to the GaAs substrate.  The Fe/{\AlGaAs} Schottky is contacted by wirebonding to a Au pad on the top of the device.  Typical device dimensions are 300~$\mu$m diameter dots (Sample A) and 400-1200~$\mu$m $\times$ 80~$\mu$m bars (Sample B).

\section{\label{sec:ExptSetup}Experimental setup}
The measurements are performed at 20~K in a split-coil magneto-optical cryostat.  The devices are biased with a current source, and the voltage drop is measured across the entire device (Schottky and n-i-p junction).

  Measurements are performed in the pure Faraday and Voigt geometries and at oblique angles up to $\pm 20^{\circ}$ from pure Voigt.  Light is collected and collimated with a 25~mm diameter lens at the cryostat window (focal length of 150~mm for Faraday and 200~mm for Voigt).  Two data collection techniques are used. For low luminescence light-level samples, such as Sample B, the collimated beam is passed through a liquid crystal variable retarder (LCVR) that can be electronically switched between $\lambda/4\text{ and }3\lambda/4$ retardation, and then through a linear polarizer.  The light is collected over a variable integration time by a CCD camera mounted on a spectrometer. 
The CCD camera collects a full EL spectrum over a 40~meV window for each setting of the LCVR ($\lambda/4\text{ and }3\lambda/4$).  The total intensity of each spectrum is integrated over the full EL line and used to define the net EL circular polarization: $P_{EL} =  (I_+-I_-)/(I_++I_-)$.  

For high luminescence light-level samples, such as Sample A, data can also be collected using a photoelastic modulator (PEM) operating at 42~kHz, a linear polarizer, a 74~mm monochromator and an avalanche photodiode (APD).  The monochromator center wavelength and exit slit are set to transmit the entire EL line.  An optical chopper ($f\approx 400$~Hz) is placed at the entrance slit of the monochromator.  The output of the APD is passed through a band-pass voltage pre-amplifier and into two digital lock-in amplifiers which are referenced to the PEM and chopper frequencies.  $P_{EL}$ is defined as $V_\text{PEM}/V_\text{chopper}$ where $V_\text{PEM}$ and $V_\text{chopper}$ are the outputs of the two lock-in amplifiers.

\subsection{\label{sec:background-spinInjection-Faraday}Faraday Geometry Electroluminescence Polarization}
\begin{figure}\begin{center}
  \includegraphics*{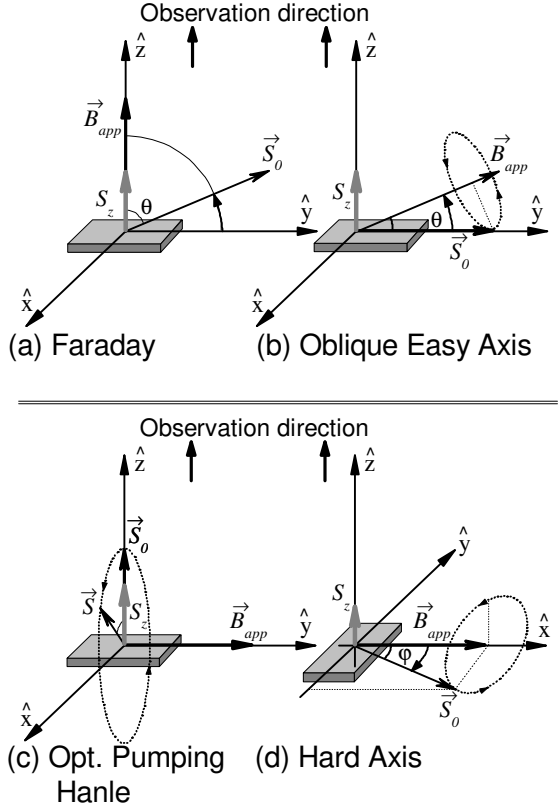}
  \caption[Experimental setup diagrams]{\label{fig:Expt-Cartoons}Measurement geometries for the experiments discussed in this paper. In all cases the observation direction is along $\hat{\bm{z}}$, the injected spin direction is along {\vso}, the applied field lies along {\vbapp}, and the detected component of steady-state spin is $S_z$.  (a) Faraday geometry: at $\mvbapp=0$, {\vso} lies entirely in-plane.  {\vbapp} is applied out-of-plane to rotate {\vso} out-of-plane. (b) Oblique easy axis geometry: {\vbapp} is applied at a small angle out-of-plane, with its in-plane projection along the [011] magnetic easy axis. {\vs} precesses about {\vbapp} after injection. (c) Voigt geometry, optical pumping Hanle effect: {\vso} is optically injected along $\hat{\bm{z}}$, and precesses about {\vbapp}, which is entirely in-plane. (d) Hard axis geometry: {\vbapp} is entirely in-plane and along the [$01\bar1$] magnetic hard axis direction. {\vso} rotates in-plane as a function of {\bapp}, and {\vs} precesses about {\vbapp} after injection.}
\end{center}\end{figure}
\begin{figure}\begin{center}
    \includegraphics*{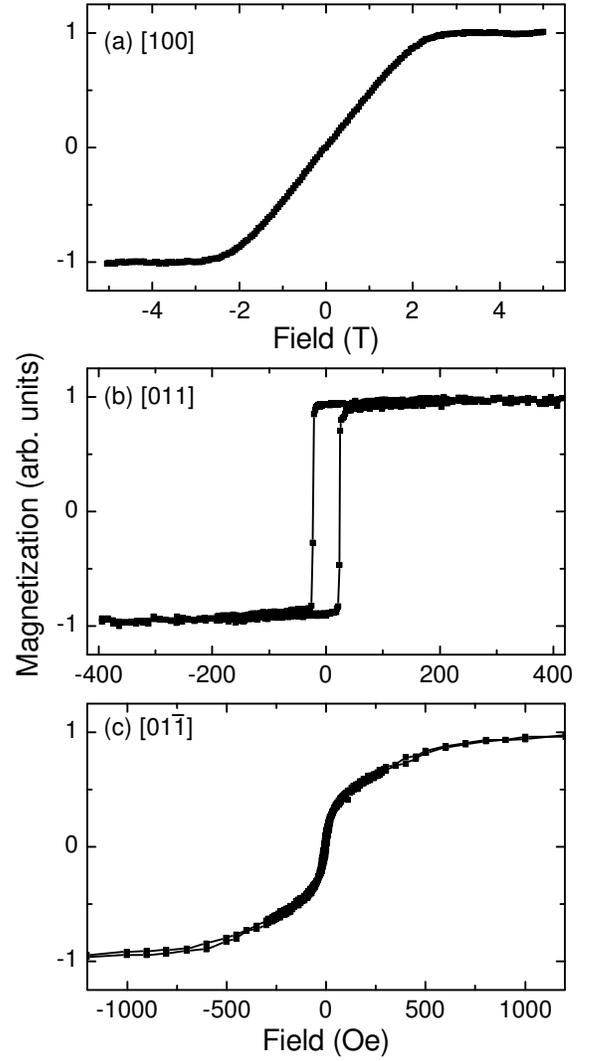}
    \caption[Fe film magnetization]{\label{fig:Magnetization}Magnetization curves for the thin Fe film.  (a) Film magnetization for magnetic field applied out-of-plane, along the [100] direction.  (b) Magnetization for magnetic field applied along the in-plane easy axis direction, [011]. (c) Magnetization for magnetic field applied along the in-plane hard axis, [$01\bar{1}$].}
\end{center}\end{figure}
\begin{figure}\begin{center}
  \includegraphics*{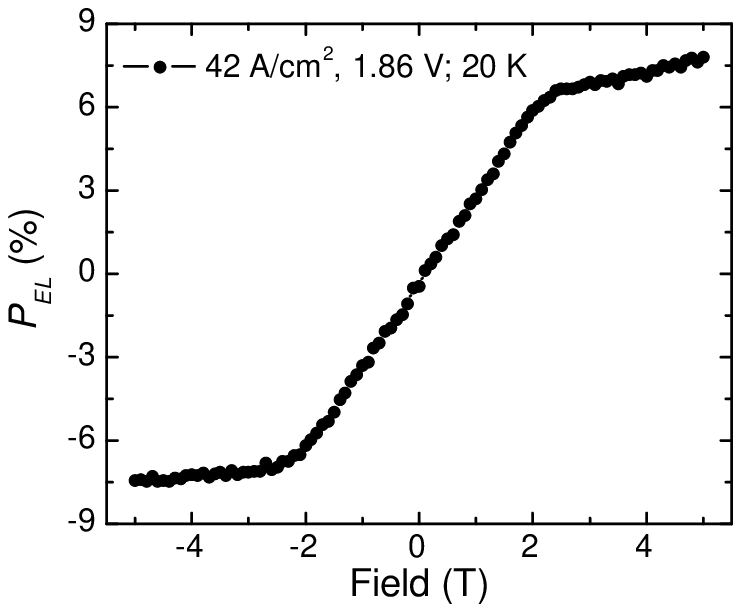}
  \caption[Faraday geometry electroluminescence polarization field sweep]{\label{fig:CA105-Faraday-FieldSweep}Electroluminescence polarization ($P_{EL}$) at fixed bias as a function of applied magnetic field along the [100] direction in the Faraday geometry. $P_{EL}$ tracks the out-of-plane component of the Fe magnetization.}
\end{center}\end{figure}
In the Faraday geometry the magnetic field is parallel to the direction of light propagation:  $\mvbapp=B_{app}\hat{\bm{z}}$ [see Fig.~\ref{fig:Expt-Cartoons}(a)]. In this case Eq.~\ref{eq:S-vector-no-Nuc} gives 
\begin{equation}
   S_z=\eta \mvso\cdot\hat{\bm{z}},
\label{eq:S-Faraday}
\end{equation}
that is, $S_z$ is simply the component of {\vso} along $\hat{\bm{z}}$ scaled by $\eta=1/(1+\tau/\tau_s)$. $\eta$ characterizes the longitudinal spin relaxation prior to recombination.  
Figure~\ref{fig:Magnetization}(a) shows the Fe film magnetization as a function of magnetic field applied out-of-plane, with the magnetization fully saturated along $\bhat{z}$ at $B_{app}=4\pi M\approx2.1$~T. Figure~\ref{fig:Expt-Cartoons}(a) shows a diagram of the Faraday geometry experimental setup: the Fe magnetization rotates out of the plane under the influence of the applied field, and EL is collected along the $z$-direction.  The increasing component of out-of-plane magnetization $M_z$ leads to an increasing $P_{EL}$ signal, until $\bm{M}$ is completely saturated out-of-plane at $\approx2.1$~T as seen in Fig.~\ref{fig:CA105-Faraday-FieldSweep}.\cite{Ploog:Zhu:first-Fe-spinLED-PRL2001, Jonker:Hanbicki:Fe-AlGaAsAPL2002, Jonker:VanTErve:AlO-Schottky2004, Jonker:Li:110-Fe-GaAs-SpinLED-APL2004, Parkin:Jiang:MTT-spinLED-PRL2003, Crowell:Adelmann-spinLED-BiasDep-PRL2004}

\begin{figure}\begin{center}
  \includegraphics*{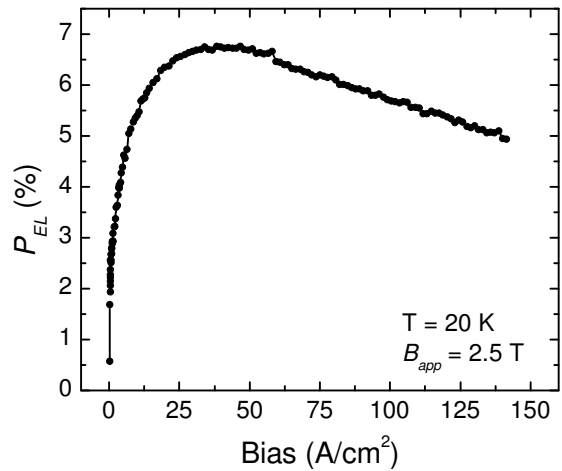}
  \caption[Faraday geometry electroluminescence polarization bias dependence]{\label{fig:CA105-Faraday-BiasDep}Electroluminescence polarization ($P_{EL}$) as a function of device bias at fixed field in the Faraday geometry. $\mbapp=2.5$~T is sufficient to saturate the magnetization along the observation direction.  The non-monotonic bias dependence of $P_{EL}$ is due to changes in the ratio of spin relaxation time to electron-hole recombination time.}
\end{center}\end{figure}

For $\mbapp>4\pi M$, $\mvso\cdot\hat{\bm{z}}=S_0$, and $S_z=\eta S_0$. Therefore, the Faraday geometry $P_{EL}$ at $B_{app}>2.1$~T directly measures the scalar value of {\etaso} after accounting for up to 1\% magneto-absorption in the semi-transparent Fe contact:
\begin{equation}\label{eq:S-Faraday-EtaS-0}
    P_{EL}=2\eta S_0.
\end{equation}
  In particular, this allows for the measurement of {\etaso} as a function of the device bias. As shown in Fig.~\ref{fig:CA105-Faraday-BiasDep}
 for $B_{app}=2.5$~T, $P_{EL}$ varies considerably and non-monotonically over the range of biases between the threshold for light emission up to the limit of the current source.  There are three physical quantities that may vary as a function of bias in the experiment of Fig.~\ref{fig:CA105-Faraday-BiasDep}: the spin injection efficiency $\epsilon$ in $S_0$, the electron-hole recombination time $\tau$, and the electron spin relaxation time $\tau_s$.  By using the combination of all three terms in the product {\etaso} for all self-consistent calculations, we do not distinguish between these three possible sources of bias dependence.  In practice, in order to extract the bias dependence of any one of the three \textit{physical} quantities it is necessary to fix the other two.  Fundamentally, however, Fig.~\ref{fig:CA105-Faraday-BiasDep} can be understood in terms of changes in the electron-hole recombination time $\tau$ and the spin relaxation time $\tau_s$, assuming essentially constant $S_0$.\cite{Crowell:Adelmann-spinLED-BiasDep-PRL2004}  At low biases, the increasing hole concentration in the QW drives $\tau$ down, increasing $\eta$.  The peak in $P_{EL}$ is associated with the window of bias in which the QW reaches ``flat bands,'' while the decrease at high bias is caused by an increase in the recombination time due to the bands bending past flat and separating the electron and hole wavefunctions.\cite{Sham:Maialle:QW-Exciton-Spin, Crowell:Adelmann-spinLED-BiasDep-PRL2004}  Using Eq.~\ref{eq:S-Faraday-EtaS-0} it is possible to relate the measured $P_{EL}$ in the Faraday geometry at any given bias, as in Fig.~\ref{fig:CA105-Faraday-BiasDep}, to a corresponding {\etaso} in Eq.~\ref{eq:S-vector-no-Nuc}.

\subsection{\label{sec:background-optical-Hanle}Voigt Geometry Photoluminescence Polarization: the Optical Hanle Effect}

A complete solution of Eq.~\ref{eq:S-vector-no-Nuc} for {\vs} requires knowledge of three sample parameters: the ratio of spin relaxation to electron recombination rates as given by $\eta$, the injected spin vector {\vso}, and the characteristic field scale for precession {\bhalf}.  The combined factor {\etaso} is measured as a function of bias in the Faraday geometry.  The direction of {\vso} is given by the magnetization of the Fe film, which can be found from the data in Fig.~\ref{fig:Magnetization} under the assumption that the magnetization rotates coherently.\cite{StonerWohlfarth}  In order to determine {\bhalf} it is necessary to perform an optical pumping Hanle effect calibration measurement.\cite{OO:DyakonovPerel}

In the Faraday geometry, the precessional motion of the injected electron spin does not affect $S_z$. In the Voigt geometry the out-of-plane component of electron spin $S_z$ is primarily a function of spin precession about {\vbapp}. The resulting change in luminescence polarization is the Hanle effect.\cite{Hanle}

Hanle curves are obtained by optically injecting spin-polarized electrons into the {\AlGaAs} barrier with circularly polarized light at an energy near the {\AlGaAs} band edge ($E_\gamma\approx1.67$~eV).  The optically injected electrons are swept into the QW and are governed by the same processes of relaxation, precession, and recombination as electrically injected carriers.  Optical pumping Hanle measurements are performed with the device bias just below the threshold for EL in order to make the conduction and valence bands as flat as possible while still minimizing background electroluminescence.  

The geometry for the optical pumping Hanle measurement is sketched in Fig.~\ref{fig:Expt-Cartoons}(c).  The magnetic field is applied entirely in-plane and the optically injected spin is along the direction of laser propagation, $\mvso^\text{\,opt}=S_0^\text{\,opt}\hat{\bm{z}}$.  Solving Eq.~\ref{eq:S-vector-no-Nuc} for this configuration gives
\begin{eqnarray}\label{eq:S-OpticalHanle}
    S_z&=&\eta\frac{S_0^\text{\,opt}B_{1/2}^2}{B_{1/2}^2+B_{app}^2},
\end{eqnarray}
which is a Lorentzian with half-width at half-maximum of {\bhalf}.  Of the four variables in Eq.~\ref{eq:S-OpticalHanle}, only two may potentially vary with bias: the coefficient $\eta$ and {\bhalf}.  In both cases, the variation with bias will be due to changes in the recombination time $\tau$, and the spin relaxation time $\tau_s$: 
 \begin{eqnarray} 
 \eta &=& \frac{1}{1+\tau/\tau_s}; \\ 
 B_{1/2}&=&\frac{\hbar}{\mgfctrEff\mu_B}\left (\frac{1}{\tau}+\frac{1}{\tau_s}\right ).  \end{eqnarray} 
However, as stated in the previous section, the bias dependence will be primarily driven by changes in the recombination time $\tau$.  The recombination time is sensitive to both the bias dependent population of holes in the QW and the formation of excitons.\cite{Crowell:Adelmann-spinLED-BiasDep-PRL2004, Sham:Maialle:QW-Exciton-Spin, Flatte:Lau:spin-relaxation-PRB2001, Flatte:Lau:DP-spin-relaxation-longCondMat2004}
For the samples under consideration here, we find $\tau\sim4\tau_s$, and $\eta$ will be more strongly dependent on changes in $\tau$ than {\bhalf}.  This is confirmed by observing negligible change in {\bhalf} over a full bias range for optical Hanle curves on a spin-LED device with very low electroluminescence intensity.  Hence, it is possible to extract a fixed value of {\bhalf} from optical pumping Hanle curves collected below EL threshold, and apply that value to calculations of {\vs} under electrical bias.  At 20~K for Sample A, $B_{1/2}=2.4$~kG.  Hence, it is possible to solve Eq.~\ref{eq:S-vector-no-Nuc} for the electrically injected steady-state spin {\vs} in the GaAs QW by combining {\bhalf} from optical pumping Hanle curves, {\etaso} as a function of bias found in Faraday geometry $P_{EL}$, and the direction of {\vso} taken from the measured Fe magnetization.

\subsection{\label{sec:background-spinInjection-Hanle}Voigt Geometry Electroluminescence Polarization: the Oblique Easy Axis Configuration}

Figure~\ref{fig:Expt-Cartoons}(b) describes the first type of Voigt $P_{EL}$ measurement discussed in this paper.  The magnetic field is aligned along the [011] direction of the sample, which is the magnetic easy axis of the Fe contact.  The field is then rotated out of the plane about the [$01\bar{1}$] axis by an angle $\theta$, $-20^\circ<\theta<+20^\circ$.  We will refer to this configuration as the \textit{oblique easy axis geometry} since the projection of {\vbapp} onto the sample plane continues to lie along the easy axis, even as $\theta$ is varied.\cite{IMEC:Motsnyi:OHE-APL2002, IMEC:motsnyi:PRB2003, IMEC:VanDorpe:AlO-Schottky2004}  The magnetization reversal along the magnetic easy axis is shown in Fig.~\ref{fig:Magnetization}(b) and is just that of a simple square hysteresis loop. The solution for $S_z$ in this setup can be found from Eq.~\ref{eq:S-vector-no-Nuc} by setting $\mvbapp=B_{app}[\cos(\theta)\hat{\bm{y}}+\sin(\theta)\hat{\bm{z}}]$, and $\mvso=S_0^h\hat{\bm{y}}$ where $S_0^h=\pm S_0$ corresponding to the two branches of the magnetic hysteresis loop:
\begin{eqnarray}
    \label{eq:S-EasyAxisVoigt}
    S_z&=&\eta\frac{S_0^h\cos(\theta)\sin(\theta)}{1+\left(B_{1/2}/B_{app}\right)^2}. 
\end{eqnarray}

\begin{figure}\begin{center}
    \includegraphics*{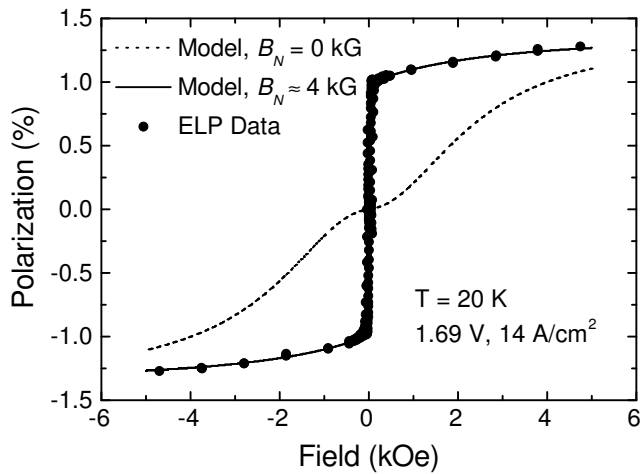}
    \caption[Oblique easy axis electroluminescence polarization curve]{\label{fig:EasyStepwModel}Electroluminescence polarization curve in the oblique easy axis geometry with $\theta=20^\circ$. The points are data, the dashed line is a calculation of the expected luminescence polarization based on the bare applied field.
The solid curve is the expected polarization including an effective magnetic field due to spin-polarized nuclei.}
\end{center}\end{figure}
Electroluminescence polarization data obtained in the oblique easy axis geometry at $\theta=20^\circ$ are shown in Fig.~\ref{fig:EasyStepwModel}.  
The points are data and the dashed line is the result of calculating $S_z$ using Eq.~\ref{eq:S-EasyAxisVoigt} with {\etaso} determined from the Faraday geometry measurements in Fig.~\ref{fig:CA105-Faraday-BiasDep}, {\bhalf} determined from an optical pumping Hanle curve, and the direction of {\vso} set by the easy axis magnetization in Fig.~\ref{fig:Magnetization}(b).  Clearly, the data does not match the $S_z$ calculation as described by Eq.~\ref{eq:S-EasyAxisVoigt}: the electron spin precesses faster in small field (further out-of-plane) then expected for the effective $\mathrm{g}$-factor in the QW $\mathrm{g}^*\approx-0.21$, leading to a large step across zero.\cite{Malinowsky:Harley-AlGaAs-GaAsQW-gFactor-PRB2000, en:eff-g-fctr}  To match the observed $P_{EL}$ it is necessary to include an effective magnetic field of $\sim$4~kG in addition to the applied magnetic field.  The presence of an effective magnetic field that is significantly larger than the applied field can be seen in the step across zero and the near saturation of $P_{EL}$ as would be expected from Eq.~\ref{eq:S-EasyAxisVoigt} for $B_{app}\gg B_{1/2}$.  The origin of the effective field is hyperfine interactions between the spin-polarized electrons and lattice nuclei in the QW that lead to dynamic nuclear polarization.  

\subsection{\label{sec:background-DNP}Dynamic Nuclear Polarization}

Dynamic nuclear polarization occurs in a solid when a non-equilibrium electron spin polarization is transferred to the nuclear spin system via the hyperfine interaction.\cite{Overhauser, Lampel:First-opt-DNP, Abragam, OO}  Dynamic nuclear polarization can be driven by saturation of an electron spin resonance,\cite{Overhauser} a large splitting between electron spin states in an applied magnetic field,\cite{en:Clark-Feher, Clark-Feher:DNP-in-InSb-byDC:PRL1963} or by injection of a non-equilibrium electron spin polarization.  There are several mechanisms for injecting the non-equilibrium electron spin polarization into a semiconductor heterostructure.  The most common mechanism is optical injection,\cite{OO} particularly for bulk GaAs,\cite{Paget:GaAsNuclearSpinCoupling-PRB1977, Paget:GaAs-donor-relaxation-PRB1982, OO} but also for quantum wells\cite{Flinn:Kerr:1stOptNMR_single_GaAsQW, Ohno:Sanada:DNP-in-110-QW-PRB2003, Awsch:Poggio:DNP-Gated-NMR-PRL2003, Awsch:Salis:OptNMR-PRB2001} and quantum dots.\cite{Gammon:QD_hyperfine_interaction} A more recent technique for generating the non-equilibrium electron spin polarization is by optical pumping at the interface between a ferromagnetic material and GaAs.\cite{Awsch:Kawakami:FM-imprinting-in-GaAs, Awsch:Epstein:FPP-PRB2002, Awsch:Epstein:FPP-in-nGaAs-PRB2002, Awsch:Epstein:FPP-VoltageControl-PRB2003, Awsch:Epstein:FPP-voltControl-PRB2003, Awsch:Stephens:Spatial-DNP-by-FPP-PRB2003, Awsch:Stephens:Spatial-DNP-Imaging-PRB2003, Awsch:Stephens:DNP-doughnuts-APL2004} Spin-blockade techniques have also been used recently to observe hyperfine interactions in GaAs  quantums dots.\cite{Tarucha:Ono:DNP-by-spinBlockade-QD-PRL2004} This report discusses a particularly simple mechanism for generating DNP in a GaAs-based QW heterostructure, direct electrical injection of spin-polarized carriers from a ferromagnetic metal across a tunnel barrier.\cite{Crowell:Strand-spinLED-DNP-PRL2003, Crowell:Strand-spinLED-DNP-NMR-APL2003, IMEC:VanDorpe:AlO-Schottky2004}  The Hamiltonian for the hyperfine interaction can be written as 
\begin{equation}
\mathcal{H}=-\frac{16\pi}{3I}\mu_B\mu_n|\Psi(R)|^2{\bhat I}\cdot{\bhat S},
\label{eq:HyperfineHamiltonian}
\end{equation}
where ${\bm I}$ is the nuclear spin, ${\bm S}$ is the electron spin, $|\Psi(R)|^2$ is the probability density of the electron wavefunction at the position of the nucleus, $\mu_n$ is the nuclear magnetic moment and $\mu_B$ is the Bohr magneton.\cite{OO, Paget:GaAsNuclearSpinCoupling-PRB1977}  Dynamic nuclear polarization occurs when spin-polarized electrons and lattice nuclei engage in hyperfine ``flip-flop'' interactions 
in which they exchange angular momentum.   For hyperfine interactions in an applied magnetic field, the difference in Zeeman energies for electrons and nuclei requires that the spin flip-flop be accompanied by an assisting process to conserve energy, such as absorption or emission of phonons or photons. A non-equilibrium electron spin polarization can therefore generate a non-equilibrium nuclear polarization at a rate $T_{pol}^{-1}$ determined by the strength of the hyperfine interaction and the rate of the assisting process:
\begin{equation}\label{eq:tpol}
    T_{pol}^{-1}=\Lambda\tau_{pol}^{-1},
\end{equation}
where $\Lambda$ contains the matrix element for the hyperfine interaction and $\tau_{pol}^{-1}$ is the rate of the assisting process that conserves energy.\cite{en:Gammon-site}  
A typical nuclear polarization time constant {\tpol} is on the order of 10 seconds in GaAs.\cite{OO:DyakonovPerel, Paget:GaAsNuclearSpinCoupling-PRB1977}

Whereas nuclear polarization is due almost exclusively to hyperfine flip-flop interactions, nuclear depolarization involves a combination of hyperfine and nuclear spin-spin interactions.  Nuclear spin relaxation is driven by precession in the fluctuating local magnetic field due to dipole-dipole interactions with neighboring nuclei $B_L\approx1.45$~G,\cite{Paget:GaAsNuclearSpinCoupling-PRB1977} and by precession about the hyperfine field of the electron spin.  The Zeeman splitting of the nuclear spin sub-levels increases the energy required to induce a transition, reducing the rate of spin relaxation by the ratio of the square of the Zeeman energies of the local and applied fields: $B_L^2/B_{app}^2$.\cite{Abragam, OO:FleischerMerkulov} The combination of the hyperfine interaction, spin-spin interactions, and Zeeman splitting can be represented for nuclear spin depolarization in a form similar to Eq.~\ref{eq:tpol}:
\begin{equation}\label{eq:T1}
    T_1^{-1}=\Lambda\tau_{depol}^{-1}\left(\frac{B_L^2}{B_{app}^2}\right),
\end{equation}
where $\Lambda$ again contains the matrix element for the hyperfine interaction, and $\tau_{depol}^{-1}$ is the rate of the assisting relaxation processes.
Because of the ten orders of magnitude difference between the spin relaxation rate of electrons ($\tau_s\sim 10^{-9}$~sec) and nuclei ($T_1\gtrsim 1$~sec), even a very weak hyperfine interaction and small steady-state electron spin polarization of a few percent can produce significant nuclear spin polarization.\cite{OO:DyakonovPerel}

A phenomenological rate equation for dynamic nuclear polarization by spin-polarized electrons can be constructed from the balance of nuclear polarization and depolarization rates $T_{pol}^{-1}$ and $T_1^{-1}$:\cite{Abragam}
\begin{eqnarray}
\frac{d\langle I_z\rangle}{dt} &=& -\frac{1}{T_{pol}}\left[\langle I_z\rangle-k\langle S_z\rangle \right]-\frac{1}{T_1}\langle I_z\rangle, \label{eq:NuclearRate}; \\
k&=&\mfl\frac{I(I+1)}{s(s+1)},
\end{eqnarray}
where $\langle S_z\rangle$ and $\langle I_z\rangle$ are the electron and nuclear spin along the applied magnetic field ${\bm B}_{app}=B_{app}\hat{\bm{z}}$, and {\fl} is a leakage factor.  A rate equation analogous to Eq.~\ref{eq:NuclearRate} can also be derived from a rigorous spin temperature argument,\cite{Abragam} in which case {\fl} is shown to be the fraction of nuclear spin relaxation that is due to hyperfine interactions.  If nuclei only relax via electrons, then $\mfl=1$, while the presence of any other relaxation mechanisms will reduce {\fl}.  Equation~\ref{eq:NuclearRate} can be solved in steady-state to find the average nuclear spin $\bm{I}_{av}=\langle I_z \rangle\hat{\bm{z}}$ along the magnetic field in terms of the electron spin and the polarization and depolarization rates:
\begin{equation}
{\bm I}_{av} = k \frac { \langle S_z\rangle \hat{\bm{z}}} {1 + \left(\frac{T_{pol}}{T_1}\right) }.
 \label{eq:I(Tpol-T1)}
\end{equation}
The hyperfine interaction results in nuclear spin aligned along the electron spin direction. The nuclei, however, precess about the applied field just as the electrons do.  While the electron spin lifetime is short, limiting precession to less than a full cycle for the field scales investigated here, the nuclear spin lifetime is long, resulting in an averaging of nuclear spin components perpendicular to {\vbapp}.
The magnetic field dependence of the average nuclear spin polarization is found by substituting for $T_{pol}$ and $T_1$ in Eq.~\ref{eq:I(Tpol-T1)}, giving
\begin{eqnarray}
{\bm I}_{av}&=& k \frac { ({\bm S} \cdot \mvbapp) \mvbapp} {B_{app}^2 + \xi B_{L}^2 },
 \label{eq:I(B)}
\end{eqnarray}
where we have used
\begin{eqnarray}
\frac{T_{pol}}{T_1} &=& \xi{\left ( {\frac{B_L}{B_{app}}}\right )}^2;\label{eq:XiBL-to-rates}\\
\label{eq:xi}
\xi &=&\frac{\tau_{pol}}{\tau_{depol}}.
\end{eqnarray}
The coefficient $\xi$ characterizes the assisting processes that enable nuclear spin polarization and depolarization but are not explicitly related to the hyperfine interaction matrix element. \\ 

The average effect of the hyperfine interactions between an electron and all nuclei within the electron's wavefunction is an effective magnetic field 
\begin{eqnarray}\label{eq:BN=bNIav}
\mvBn &=& \sum_{\alpha} b_N^\alpha {\bm I}_{av}^\alpha/I^{\alpha};\\
b_N^\alpha &=& \frac{16\pi}{3\mathrm{g}^*v_0}\mu_n^\alpha d_e^\alpha N^\alpha,
\end{eqnarray}
where $\mathrm{g}^*$ is the effective electron $\mathrm{g}$-factor, $v_0$ is the volume of the unit cell, $d_e^\alpha$ is the electron wavefunction density at the $\alpha$ isotope nucleus, $\mu_n^\alpha$ is the nuclear magnetic moment, and $N^\alpha$ is the number of $\alpha$ nuclei in the unit cell.\cite{Paget:GaAsNuclearSpinCoupling-PRB1977, OO:DyakonovPerel}  Given that all three isotopes in GaAs ($^{75}\text{As}$, $^{69}\text{Ga}$, and $^{71}\text{Ga}$) have the same spin $I=3/2$, we replace $I^{\alpha}$ with $I$.  Additionally, since we are unable to distinguish the contributions of individual isotopes to the total effective field, we replace $b_N^\alpha$ with a total {\bn} that includes the contributions of all nuclei.  This {\bn} is the maximum effective magnetic field felt by electrons for 100\% spin-polarized nuclei.  In bulk GaAs with $\mathrm{g}^*_\text{bulk}=-0.44$, $b_N^\text{bulk}=53$~kG.\cite{Paget:GaAsNuclearSpinCoupling-PRB1977, OO:DyakonovPerel}  In a {100~\AA} {\AlGaAs/GaAs} QW, $b_N^\text{QW}=\left(g_\text{bulk}^*/g_\text{QW}^*\right)b_N^\text{bulk}\approx 111$~kG.  For the remainder of the paper, we will use $b_N=111$~kG.  Combining Eqs.~\ref{eq:I(B)} and \ref{eq:BN=bNIav} gives the effective magnetic field due to polarized nuclei,\cite{OO}
\begin{equation}
\mvBn=\mfl b_N\frac{(I+1)}{s(s+1)} \frac{({\bm S} \cdot \mvbapp){\bm
B}_{app}}{B_{app}^2 + \xi B_{L}^2 }.
 \label{eq:B-N}
\end{equation}

Spin-polarized electrons precess as though in the presence of a total magnetic field $\mva=\mvbapp+\mvBn$.  Equation~\ref{eq:S-vector-no-Nuc} for electron spin dynamics in the QW then becomes
\begin{equation}
   {\bm S}=\eta\frac{B_{1/2}^2{\bm S}_0+({\bm S}_0 \cdot \mva)\mva+B_{1/2}(\mva\times{\bm S}_0)}{B_{1/2}^2+A^2},
\label{eq:S-vector-Nuc}
\end{equation}
and Eq.~\ref{eq:S-EasyAxisVoigt} for the special case of the oblique easy axis configuration as in Figs.~\ref{fig:Expt-Cartoons}(b) and \ref{fig:EasyStepwModel} becomes
\begin{equation}\label{eq:S-EasyAxisVoigt-Nuc}
    S_z=\eta\frac{S_0\cos(\theta)\sin(\theta)}{1+\left(B_{1/2}^2/A^2\right)}.
\end{equation}
Using Eq.~\ref{eq:B-N} for {\Bn}, Eq.~\ref{eq:S-EasyAxisVoigt-Nuc} gives the solid line in Fig.~\ref{fig:EasyStepwModel} as a fit to the data using the leakage factor {\fl} as the only free parameter. For the solid line in Fig.~\ref{fig:EasyStepwModel}, $\mfl=0.46$.  The factor {\xibl} in Eq.~\ref{eq:B-N} is less than 50~G, and can be ignored when fitting data over wide field ranges on the order of several kOe. Compared with $B_{1/2}=2.4$~kG, it is clear how the effective field $B_N\approx 4$~kG dominates the electron spin dynamics in Fig.~\ref{fig:EasyStepwModel}, particularly at low {\bapp}.  The range of {\bapp} in which  $P_{EL}$ is most sensitive to nuclear polarization effects lies between the local field factor {\xibl} and {\bhalf}.  When $B_{app}^2\lesssim \mxiblsq$, {\Bn} approaches zero and does not significantly affect electron spin precession.  At $B_{app}^2>B_{1/2}^2$ the applied field is sufficient to produce significant electron precession and the sensitivity to nuclear polarization effects is reduced.\\

The local field factor {\xiblsq} in the denominator of Eqs.~\ref{eq:I(B)} and \ref{eq:B-N} defines the scale of Zeeman energy below which the average nuclear spin polarization goes to zero.  As discussed above, the local field $B_L\approx 1.45$~G is the effective magnetic field due to nuclear spin-spin interactions, and $\xi$ is a factor that incorporates sample-specific processes that assist nuclear spin transitions. The effect of the local field at low {\bapp} is a reduction in {\Bn} and consequently less electron spin precession.  
\begin{figure}\begin{center}
    \includegraphics*{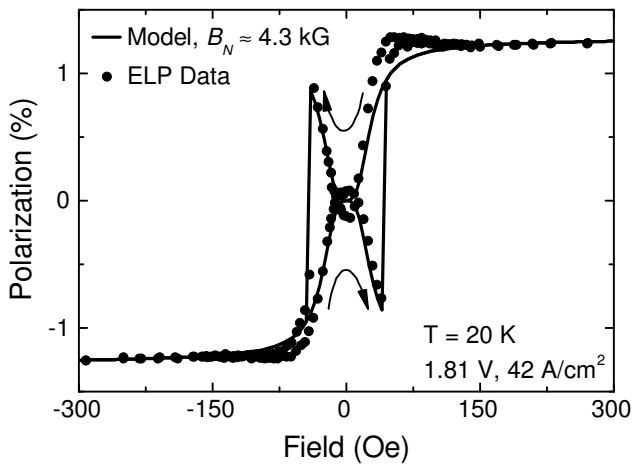}
    \caption[Oblique easy axis electroluminescence polarization hysteresis loop]{\label{fig:CA105-EasyAxis-LoopWmodel}Electroluminescence polarization hysteresis loop in the oblique easy axis geometry with $\theta=20^\circ$.  The points are data and the line is a fit which includes the effect of dynamically polarized nuclei.  The broad dips in $P_{EL}$ as the field is swept through zero are due to nuclear spin depolarization for $\mbapp^2<\mxiblsq$.}
\end{center}\end{figure}
Without the reduction in {\Bn} and spin precession at $B_{app}^2\lesssim \mxiblsq$, the oblique easy axis curve in Fig.~\ref{fig:CA105-EasyAxis-LoopWmodel} would exactly match the shape of the easy axis hysteresis loop in Fig.~\ref{fig:Magnetization}(b), with a singularity at $B_{app} = 0$~T. Instead broad dips in the luminescence polarization appear as the field is swept through zero. The width of this depolarization region can be fit by setting $B_L=1.45$~G and varying $\xi$. This is a single parameter fit since {\fl} is separately determined over a wider field scale on which {\xiblsq} has a negligible affect.  For the data in Fig.~\ref{fig:CA105-EasyAxis-LoopWmodel}, the fit gives $\xi=400\pm120$.  The uncertainty in the fit comes from the overshoot in $P_{EL}$ at $B_{app}=0$~Oe, which is attributed to small off-axis components of the field profile of the superconducting magnet as the field is swept through zero.

\section{\label{sec:Results}Experimental Results and Discussion}
\subsection{\label{sec:Results-EasyAxis}Field Dependence along the Easy Axis}
\begin{figure}\begin{center}
    \includegraphics*{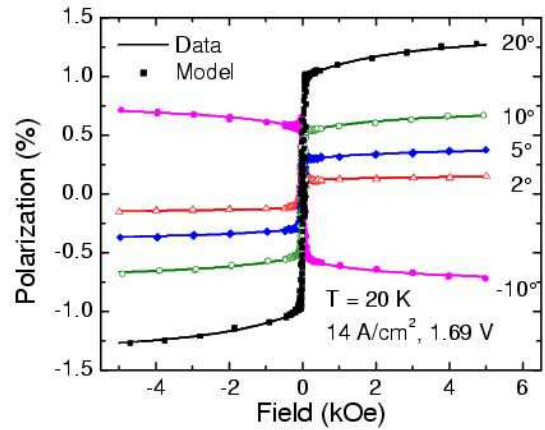}
    \caption[Oblique easy axis electroluminescence polarization curves at multiple angles]{\label{fig:EasyAngleFits-20K-CA105}Electroluminescence polarization curves in the oblique easy axis geometry at several values of $\theta$.  The points are data.  The lines are predicted polarization curves with parameters fixed by fitting the {20\degrees} data and then changing only the measurement angle for subsequent calculations.}
\end{center}\end{figure}
The large step across zero field in the single $\pm5$~kG oblique easy axis field sweep at $\theta=20^\circ$ in Fig.~\ref{fig:EasyStepwModel} was fit using Eqs.~\ref{eq:B-N} and \ref{eq:S-EasyAxisVoigt-Nuc} and a single free parameter {\fl}.  It was assumed that {\etaso} measured in the Faraday geometry and {\bhalf} measured in an optical pumping Hanle curve remained unchanged. The consistency of these assumptions can be tested by taking the parameters from Fig.~\ref{fig:EasyStepwModel} and applying them to a set of oblique easy axis curves collected with {\vbapp} at different out-of-plane angles, adjusting only $\theta$ in Eqs.~\ref{eq:B-N} and \ref{eq:S-EasyAxisVoigt-Nuc} to fit the additional curves.  This is shown in Fig.~\ref{fig:EasyAngleFits-20K-CA105}: the points are data, the line for $\theta=20^\circ$ is a one parameter fit, and the remaining lines are calculated without adjusting any parameters.  The model accurately predicts the oblique easy axis curve at each subsequent measurement angle within the experimental uncertainty $\Delta\theta\sim0.4^\circ$, confirming the validity of the fitting assumptions.

\begin{figure}\begin{center}
    \includegraphics*{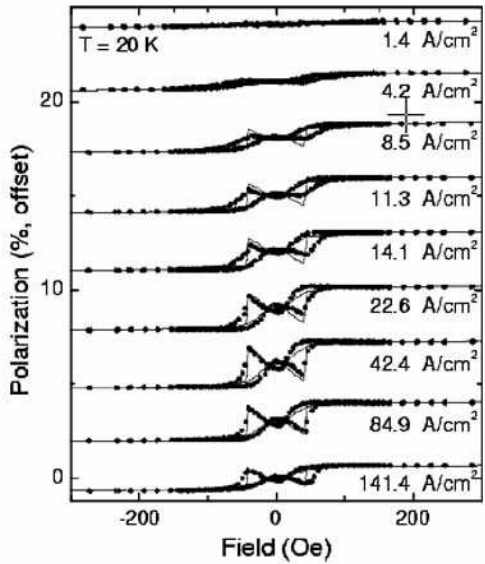}
    \caption[Oblique easy axis electroluminescence polarization hysteresis loops as a function of bias]{\label{fig:CA105-EasyAxis-BiasDep}Electroluminescence polarization ($P_{EL}$) hysteresis loops in the oblique easy axis geometry with $\theta=20^\circ$ at several biases.  The points are data and the lines are fits which include the effect of dynamically polarized nuclei.  The amplitude of the oblique easy axis curves varies with device bias due to changes in the steady-state electron spin polarization.  The $P_{EL}$ loops do not fully close at the coercive field for the Fe contact because the nuclear polarization lags as {\bapp} is swept at a rate comparable to $T_1^{-1}$. All curves are offset for clarity.}
\end{center}\end{figure}
Figure~\ref{fig:CA105-EasyAxis-BiasDep} contains several oblique easy axis curves at increasing bias over a smaller field range than the curves in Fig.~\ref{fig:EasyAngleFits-20K-CA105}. Over the smaller scale, the Fe contact hysteresis can be observed and the {\xiblsq} factor in the denominator of Eq.~\ref{eq:B-N} becomes relevant.  The points in Fig.~\ref{fig:CA105-EasyAxis-BiasDep} are data, and the lines are fits in which {\etaso} is taken from Faraday geometry measurements at the corresponding bias, and {\bhalf} is taken from the optical pumping Hanle curve. The parameter $\xi=520$, and thus $\mxiblsq=1100$~G$^2$, is fixed for all curves by fitting the low field oblique easy axis $P_{EL}$ depolarization features in Fig.~\ref{fig:CA105-EasyAxis-LoopWmodel}. {\fl} is the only remaining free parameter that is adjusted to fit the data at each bias. All curves are offset for clarity.  The $P_{EL}$ amplitude in Fig.~\ref{fig:CA105-EasyAxis-BiasDep} is defined as the difference in $P_{EL}$  between $+100$~Oe and $-100$~Oe. At low biases, just above the threshold for light emission, the oblique easy axis curve closely resembles the prediction for  zero nuclear polarization. The $P_{EL}$ amplitude increases as the bias initially increases above threshold, reaches a maximum, and then decreases at higher biases. The fits shown in Fig.~\ref{fig:CA105-EasyAxis-BiasDep} accurately capture the amplitude change as a function of bias.  There are, however, discrepancies between the $P_{EL}$ data and the calculated hysteresis loop shape in Fig.~\ref{fig:CA105-EasyAxis-BiasDep}: the data does not completely close at the coercive field for the Fe contact as predicted by the model.  This discrepancy is due to a lag in the response of {\Bn} to changes in {\bapp}, which is being swept at a rate comparable to $T_1^{-1}$.  The calculation gives the steady-state value of {\Bn} and $S_z$ at each point. Oblique easy axis field sweeps taken with slower field sweep rates more closely match the square switching features of the calculated loops.  For example, for the data run shown as the oblique easy axis loop in Fig.~\ref{fig:CA105-EasyAxis-LoopWmodel}, there was a 20~second wait time between setting {\bapp} and measuring $P_{EL}$ for each point, and the data more closely matches the steady-state model.

The amplitude change observed in oblique easy axis hysteresis loops as a function of bias is directly linked to the changing effective magnetic field {\Bn}: initially increasing with bias, then peaking and decreasing.  The changes in {\Bn} are in turn related to changes in steady-state electron spin polarization in the QW as measured in the Faraday geometry.
\begin{figure}\begin{center}
  \includegraphics*{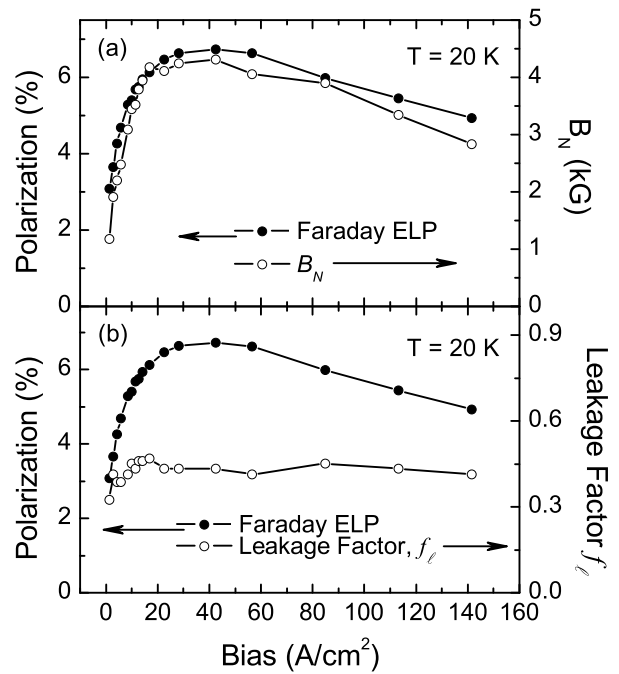}
  \caption[Effective magnetic field and nuclear polarization leakage factor as a function of bias]{\label{fig:CA105-EasyAxis-20K-B-N-FaradayELP}Both panels, left axis (closed symbols): steady-state electron spin polarization in the QW as measured by Faraday geometry $P_{EL}$ at 2.5~T.  (a) Right axis (open symbols): the asymptotic value of the effective magnetic field {\Bn} due to spin-polarized nuclei based on fits to oblique easy axis geometry $P_{EL}$ curves. (b) Right axis (open symbols): values of the leakage factor {\fl} used to determine {\Bn} in (a). }
\end{center}\end{figure}
The closed symbols in Fig.~\ref{fig:CA105-EasyAxis-20K-B-N-FaradayELP} plot Faraday $P_{EL}$ at $B_{app}=2.5$~T (left axis).  This signal is the steady-state electron spin polarization {2\etaso}, and is used to fit the oblique easy axis curves as discussed in Section~\ref{sec:background-spinInjection-Faraday}.  The open symbols (right axis) in Fig.~\ref{fig:CA105-EasyAxis-20K-B-N-FaradayELP}(a) are the asymptotic values of {\Bn} at high field from fits to curves such as those in Fig.~\ref{fig:CA105-EasyAxis-BiasDep}.  The nearly identical qualitative bias dependence of both {\Bn} and the Faraday geometry $P_{EL}$ reflects the role of {\vs} in Eq.~\ref{eq:B-N}, and the close coupling between steady-state electron spin polarization in the QW and nuclear polarization. The other two terms in Eq.~\ref{eq:B-N} that may vary with bias are {\fl} and $\xi$.  The leakage factor {\fl} is the one free fitting parameter for the data in Fig.~\ref{fig:CA105-EasyAxis-BiasDep} and is plotted as a function of device bias in Fig.~\ref{fig:CA105-EasyAxis-20K-B-N-FaradayELP}(b) (right axis, open symbols).  {\fl} increases slightly at low bias, then remains essentially unchanged at $\sim0.41$ throughout the rest of the operational bias range, reflecting a lack of significant variation in nuclear spin relaxation mechanisms as a function of bias above threshold.  This result is somewhat surprising. One would expect the balance of nuclear depolarization mechanisms to vary as a function of electron density in the QW. However this does not appear to be the case. Instead, the nuclear system responds primarily to changes in the electron spin polarization $S$.

\subsection{\label{sec:Results-HardAxis}Field Dependence along the Hard Axis}

\begin{figure} \begin{center}
    \includegraphics*{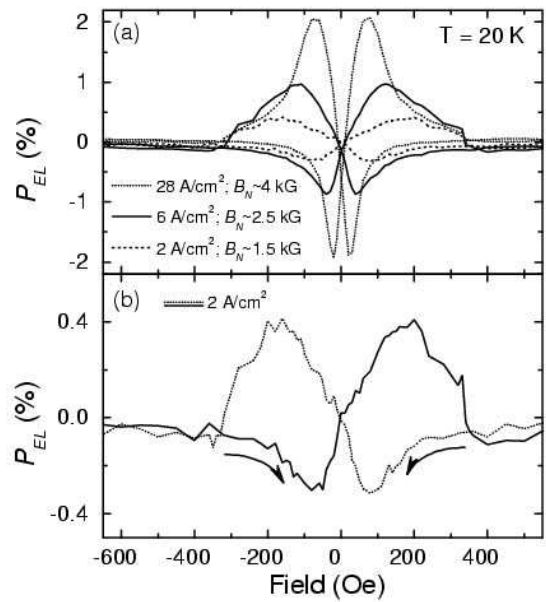}
    \caption[Hard axis electroluminescence polarization hysteresis loops as a function of bias]{\label{fig:CA105-HardAxis-LoopsData}Electroluminescence polarization hysteresis loops in the hard axis geometry.  (a) Hard axis $P_{EL}$ loops at three biases, and therefore three different effective magnetic fields {\Bn}.  (b) Magnified view ($\times5$) of the lowest bias hard axis $P_{EL}$ loop from (a). The dotted line indicates the signal for {\bapp} sweeping from positive to negative field. The solid line indicates {\bapp} sweeping from negative to positive field.}
\end{center} \end{figure}

  In order to detect an out-of-plane component of spin in the case of {\vbapp} near the magnetic easy axis ([011]), the field must be rotated out of the plane to create an angle between {\vso} and {\vbapp}. For fields applied along the magnetic hard axis ([$01{\bar 1}$]), there is no need to rotate the field out of the plane since the magnetization itself will coherently rotate through a full in-plane circle as {\vbapp} is swept from positive to negative values and back.  We will refer to this configuration as the \textit{hard axis geometry}.  The assumption of magnetization reversal by coherent rotation\cite{StonerWohlfarth} is confirmed by fitting in-plane magnetization curves with the field along the hard axis $[01\bar{1}]$ and at {45\degrees} between the hard and easy axes, along [010].  As can be seen from the hard axis magnetization curve in Fig.~\ref{fig:Magnetization}(c), for $B_{app}>500$~Oe, {\mhat} is nearly parallel to $\mvbapp$.  As the field decreases, {\mhat} rotates away from [$01{\bar 1}$] and towards the [011] easy axis. The rotation direction is determined by any slight in-plane misalignment between [$01{\bar 1}$] and the field axis. The diagram in Fig.~\ref{fig:Expt-Cartoons}(d) and the following discussion assumes {\mhat} rotates from $+\hat{\bm{x}}$ towards $-\hat{\bm{y}}$. As {\vbapp} goes through zero, {\mhat} rotates through the $-\hat{\bm{y}}$ easy axis, and saturates along the hard axis in the $-\hat{\bm{x}}$ direction. As {\vbapp} sweeps back to complete the hysteresis loop, {\mhat} rotates through the $+\hat{\bm{y}}$ easy axis and realigns with the $+\hat{\bm{x}}$ hard axis completing the full circle.  Setting $\mvbapp=B_{app}\hat{\bm{x}}$ and $\mvso=S_0[\cos(\phi)\hat{\bm{x}}+\sin(\phi)\hat{\bm{y}}]$ where $\phi$ is the angle between {\mhat} and the $x$-axis, the $z$-component of steady-state electron spin from Eq.~\ref{eq:S-vector-Nuc} is
\begin{equation}\label{eq:S-HardAxis-Nuc}
    S_z=\eta\frac{S_0AB_{1/2}\sin{\phi}}{B_{1/2}^2+A^2},
\end{equation}
which results in the double-lobed hysteresis curves shown in Fig.~\ref{fig:CA105-HardAxis-LoopsData}.  
 
Figure~\ref{fig:CA105-HardAxis-LoopsData}(a) shows three electroluminescence polarization hysteresis loops in the hard axis geometry taken at different biases, and therefore different effective magnetic fields {\Bn}.  The data exhibit changes in both the peak-to-peak $P_{EL}$ amplitude and the shape of the loops as a function of increasing {\Bn}. The dotted curve at the higher bias and larger effective magnetic field is stretched vertically, tracing out tall, pointed lobes with peaks near zero field.  The solid and dashed curves at lower bias and effective magnetic field are accordingly smaller with more rounded lobes and peaks further from zero field.  Figure~\ref{fig:CA105-HardAxis-LoopsData}(b) is an expanded view ($\times5$) of the lowest bias curve (dashed line) in Fig.~\ref{fig:CA105-HardAxis-LoopsData}(a).  Figure~\ref{fig:CA105-HardAxis-LoopsData}(b) shows more clearly the broader and more widely spaced lobes that correspond to lower {\Bn}, as well as the hard axis $P_{EL}$ trace for the two field sweep directions. The dotted line is for {\bapp} sweeping from positive to negative fields, and the solid line is for {\bapp} sweeping from negative to positive fields.

\begin{figure}\begin{center}
    \includegraphics*{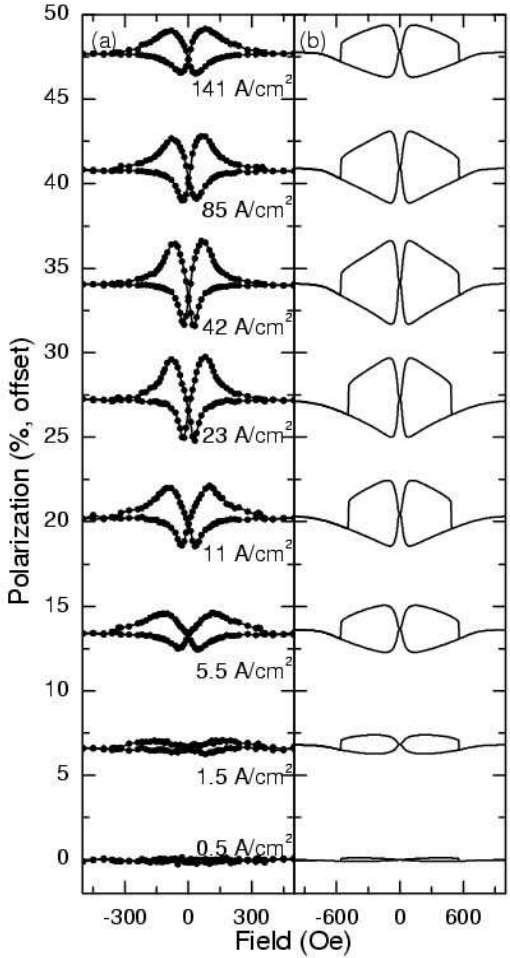}
    \caption[Hard axis electroluminescence polarization hysteresis loops as a function of bias with modeling]{\label{fig:CA105--HardAxis--NoFitting}(a) Electroluminescence polarization hysteresis loops in the hard axis geometry at several biases. (b) Predicted polarization in the hard axis geometry based on fits to the oblique easy axis geometry $P_{EL}$ curves at corresponding biases. There are no free parameters used to generate the curves in (b). All curves are offset for clarity.}
\end{center}\end{figure}

The amplitude change as a function of bias in Fig.~\ref{fig:CA105-HardAxis-LoopsData}(a) is expected in light of the strong $P_{EL}$ amplitude dependence as a function of bias observed in the easy axis geometry (Fig.~\ref{fig:CA105-EasyAxis-BiasDep}).  The shape change observed for $P_{EL}$ hysteresis loops in the hard axis geometry is predicted in Eq.~\ref{eq:S-HardAxis-Nuc} as the total internal magnetic field $A$ changes from less than {\bhalf} to greater than {\bhalf}.  Figure~\ref{fig:CA105--HardAxis--NoFitting}(a) shows $P_{EL}$ hysteresis loops at several different biases in the hard axis geometry using the same sample and bias range as the oblique easy axis loops in Fig.~\ref{fig:CA105-EasyAxis-BiasDep}---the curves are vertically offset for clarity.   
Just as in the case of the oblique easy axis geometry, the change in the effective magnetic field causes a modulation in loop amplitude that mirrors that of the steady-state electron spin polarization. The values of {\fl} and $\xi$ found in the oblique easy axis geometry (in Figs.~\ref{fig:CA105-EasyAxis-20K-B-N-FaradayELP}(b) and \ref{fig:CA105-EasyAxis-LoopWmodel}, respectively)  can be used directly to calculate $P_{EL}$ hysteresis loops in the hard axis geometry with full consistency.  This is shown in Fig.~\ref{fig:CA105--HardAxis--NoFitting}(b) with no parameters adjusted. There is good overall agreement between the data and the calculations as a function of bias with respect to both changes in the $P_{EL}$ loop amplitude and loop shape.  For the calculations in Fig.~\ref{fig:CA105--HardAxis--NoFitting}(b), the angle $\phi(\mbapp)$ between $\bm M$ and {\vbapp} is generated using a first principles Stoner-Wohlfarth model.\cite{StonerWohlfarth}  The Stoner-Wohlfarth model is in turn fit to measurements of the hard axis magnetization reversal of the continuous ($\sim 5$~mm$\times 5$~mm) 50~{\AA} Fe film before the sample is processed into individual devices. The quantitative discrepancies between the data in Fig.~\ref{fig:CA105--HardAxis--NoFitting}(a) and the calculations in Fig.~\ref{fig:CA105--HardAxis--NoFitting}(b) are due to the difference between the fit to the  magnetization reversal of the continuous film (area $\sim 25$~mm$^2$) and the actual magnetization reversal of the processed device (area $\sim 0.07$~mm$^2$).

\subsection{\label{sec:Results-timedep}Time Dependence of the Nuclear Polarization }

\begin{figure*}
    \includegraphics*{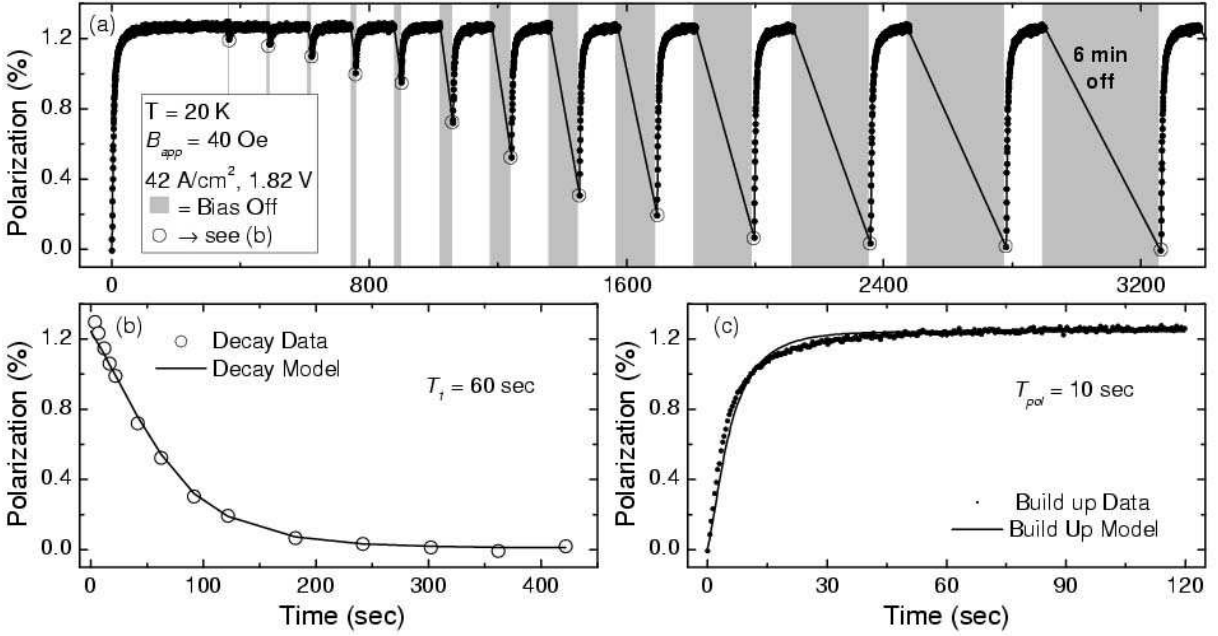}
    \caption[Electroluminescence polarization as a function of time]{\label{fig:CA105-TimeDep-EasyAxis+40G}Electroluminescence polarization ($P_{EL}$) as a function of time at fixed field in the oblique easy axis geometry with $\theta=20^\circ$.  (a) $P_{EL}$ as a function of laboratory time as the device bias is switched on and off.  The gray bars indicate times during which the bias is off and the nuclear polarization decays. Data cannot be collected when the bias is off.  The amount of nuclear spin polarization decay increases with increasing bias off time.  (b) Reconstructed $P_{EL}$ decay as a function of bias off time.  Points are taken from the initial (circled) point after the bias is turned on in (a), and plotted against the preceding off time.  The curve is a fit to the data assuming that the nuclear polarization decays exponentially with time constant {\tone}. (c) $P_{EL}$ as a function of time after the bias is turned on.  {\Bn~= 0} at $t=0$. The line is  a fit to the data assuming a saturating exponential polarization with time constant {\tpol}.}
\end{figure*}

To first order, nuclear spin polarization and depolarization, and hence the effective magnetic field, will be exponential in time after the spin-polarized current is turned on or off.  The depolarization is characterized by the spin-lattice relaxation time {\tone} in Eq.~\ref{eq:NuclearRate}. In the absence of any polarizing mechanism, the average nuclear spin will decay from its value at $t=0$ as
\begin{equation}\label{eq:Iav-Decay}
    {\bm I}_{av}(t)={\bm I}_{av}(0)e^{-t/T_1}.
\end{equation}
Equation~\ref{eq:Iav-Decay} is incorporated into the right hand side of Eq.~\ref{eq:B-N} as
\begin{equation}\label{eq:B-NExpDecay}
\mvBn(t)=\mvBn(0)e^{-t/T_1}
\end{equation}
to give the effective magnetic field for decaying nuclear spin polarization.
Dynamic nuclear spin polarization by oriented electrons occurs in the presence of nuclear spin-lattice relaxation, and nuclear polarization is found by integrating Eq.~\ref{eq:NuclearRate} to obtain
\begin{equation}\label{eq:Iav-BuildUp}
    {\bm I}_{av}(t)={\bm I}_{av}(\infty)\left(1-e^{-t\left(\frac{T_1+T_{pol}}{T_1T_{pol}}\right)}\right ),
\end{equation}
where ${\bm I}_{av}(\infty)$ is the steady-state value of the average nuclear spin.
Inserting this exponential envelope for nuclear spin into Eq.~\ref{eq:B-N} gives a saturating exponential for the effective field.  
Although the exponential approximation does not take into account nuclear spin diffusion, it nonetheless allows for a first order evaluation of {\tone} and {\tpol}.\cite{Paget:GaAs-donor-relaxation-PRB1982}

The time dependence of nuclear polarization build up and decay can be probed by modulating either the steady-state electron spin in the device or the applied field.  The most direct experimental approach is to switch the electron spin in the QW on and off as a function of time and observe the change in nuclear polarization by the effect that change has on $P_{EL}$.  This can be done by switching the bias current on and off.  When the device bias is off, the steady-state electron spin polarization is zero, and any nuclear spin polarization will decay away with a time constant {\tone}. When the bias is turned on, the nuclear polarization will build up as in Eq.~\ref{eq:Iav-BuildUp}. At fixed {\bapp} the change in nuclear polarization will result in a change in $P_{EL}$, which can be measured as a function of time in either the oblique easy axis geometry or the hard axis geometry. The analysis is simplest in the case of the oblique easy axis geometry.

In Fig.~\ref{fig:CA105-TimeDep-EasyAxis+40G}(a), $P_{EL}$ is plotted as a function of laboratory time.  The device starts with the bias off for several minutes.  At $t=0$ the bias is turned on for 6 minutes, then turned off for 2~seconds, turned on again for 120~seconds, then off for 6~seconds, then on again for 120~seconds and so on with increasing times during which the bias is off. In each case the device is on for 120~seconds.  The gray bars in Fig.~\ref{fig:CA105-TimeDep-EasyAxis+40G} indicate times during which the bias is off. Note that data points can only be collected when the bias is on.  The sample is positioned with in the oblique easy axis configuration with $\theta=20^\circ$ after ramping {\bapp} down from +1~kOe down to +40~Oe. The resulting data show the laboratory scale time dependence of nuclear spin polarization build up and decay.

As seen from Eq.~\ref{eq:S-EasyAxisVoigt-Nuc} and the dashed line in Fig.~\ref{fig:EasyStepwModel}, at low fields $(\mbapp\ll\mbhalf)$the out-of-plane component of steady-state spin is negligible if there is no effective magnetic field from polarized nuclei to induce significant precession.  At fixed field and bias, the change in $P_{EL}$ as a function of time is a direct measure of the change in the precession angle due to the change in nuclear spin polarization.  The initial point collected after the bias is turned on depends on the precession angle in the residual effective field that remains at the end of a bias-off cycle.  By selecting the initial point after each off cycle [the circled points in Fig.~\ref{fig:CA105-TimeDep-EasyAxis+40G}(a)], and plotting $P_{EL}$ at that point against its associated off time, it is possible to construct a plot of $P_{EL}$ versus time during which the bias is off, as shown in Fig.~\ref{fig:CA105-TimeDep-EasyAxis+40G}(b).  Since the out-of-plane component of electron spin is proportional to the nuclear polarization in this geometry, the time constant for $P_{EL}$ decay in Fig.~\ref{fig:CA105-TimeDep-EasyAxis+40G}(b) is the longitudinal nuclear spin relaxation time $T_1$. This decay time plot can be fit using Eqs.~\ref{eq:S-EasyAxisVoigt-Nuc} and \ref{eq:B-NExpDecay} to give the solid line in Fig.~\ref{fig:CA105-TimeDep-EasyAxis+40G}(b) with $T_1\approx60$~seconds.

  Figure~\ref{fig:CA105-TimeDep-EasyAxis+40G}(c) shows $P_{EL}$ as a function of time in response to turning the device bias on after all of the nuclear polarization has decayed away.  This curve can be fit using Eqs.~\ref{eq:S-EasyAxisVoigt-Nuc} and \ref{eq:Iav-BuildUp} and the value of $T_1$ extracted from Fig.~\ref{fig:CA105-TimeDep-EasyAxis+40G}(b). The result is shown as the solid line in Fig.~\ref{fig:CA105-TimeDep-EasyAxis+40G}(c), for which $T_{pol}\approx10$~seconds.  \\

Table~\ref{tbl:times-vs-field} summarizes the results from fitting {\tpol} and {\tone} at different applied fields in the oblique easy axis geometry.  At lower {\bapp} the {\tone} ({\tpol}) is shorter (longer), as expected from Eqs.~\ref{eq:XiBL-to-rates} and \ref{eq:xi}.  
Table~\ref{tbl:times-vs-field} also includes the values of $\xi$ calculated from Eq.~\ref{eq:XiBL-to-rates}.  
\begin{table}\begin{ruledtabular}
\begin{tabular}{rrrc}
Field (Oe)   &   {\tone}~(sec) &   {\tpol}~(sec) &   $\xi$\\ \hline
$-40\pm5$\;\;\;\; &$95\pm10$\;  &$9\pm2$\;\;\;   &\:\:\;\;$72\pm42$\;\:  \\
$18\pm4$\;\;\;\;  &$17\pm10$\;  &$102\pm5$\;\;\; &$1380\pm500$ \\
$40\pm5$\;\;\;\;  &$60\pm5$\:\;\;  &$10\pm2$\;\;\;  &\;\;\:$127\pm68$ \:\: \\
$200\pm5$\;\;\;\; &$440\pm10$\; &$10\pm2$\;\;\;  &\;\;\:$432\pm118$\:  \\
$300\pm5$\;\;\;\; &$600\pm20$\; &$2.3\pm1$\;\;\; &\;\;\:$164\pm82$ \:\; \\
Field Sweep  &\multicolumn{1}{c}{--}&\multicolumn{1}{c}{--}&\:\;$400\pm120$\:
\end{tabular}\end{ruledtabular}
\caption[Field dependence of $T_1$, {\tpol} and $\xi$]{\label{tbl:times-vs-field}Values of {\tone} and {\tpol} from oblique easy axis time dependence measurements, as well as the calculated local field factor $\xi$.  The bottom row contains the value of $\xi$ from fitting the low field depolarization feature of oblique easy axis field sweeps.}
\end{table} 
The calculated values of $\xi$ at different fields, and the value found from fitting the low field depolarization feature in oblique easy axis field sweeps agree within an order of magnitude. Note that the measurements that correspond to the two extreme values of $\xi$, $\mbapp=-40$~Oe and $\mbapp=+18$~Oe are each special cases.  At $B_{app}=-40$~Oe, {\vso} and $\bm I_{av}$ are antiparallel to {\vbapp}, corresponding to a non-equilibrium, negative spin temperature configuration for the polarized nuclei.  The other extreme case, $\mbapp=18$~Oe, gives the least consistent value of $\xi$ and has the highest uncertainty.  $B_{app}=18$~Oe is at a positive nuclear spin temperature, but it is well within the low field nuclear spin depolarization region. The data at $\mbapp=18$~Oe are also the only set for which $T_{pol}>T_1$.  The value of $\xi$ determined in other experiments ranges from a low of 2--3 in bulk GaAs\cite{Paget:GaAsNuclearSpinCoupling-PRB1977, OO} to a high of $\sim3\times10^5$ in quantum dots\cite{Gammon:QD_hyperfine_interaction} and [110] GaAs quantum wells.\cite{Ohno:Sanada:DNP-in-110-QW-PRB2003} A large value of $\xi$ on the order of $5\times10^4$ is also observed in DNP driven by imprinting at a ferromagnet-semiconductor interface.\cite{Awsch:Kawakami:FM-imprinting-in-GaAs} The large variation in the observed value of $\xi$ reflects the role of the assisting processes that conserve energy during the hyperfine interaction (see Eq.~\ref{eq:xi} and Ref.~\onlinecite{Gammon:QD_hyperfine_interaction}).  The relative rates of the assisting processes will vary depending on the sample structure, resulting in dramatically different magnetic field scales for nuclear spin depolarization.

\begin{figure}
    \includegraphics*[width=8.2cm]{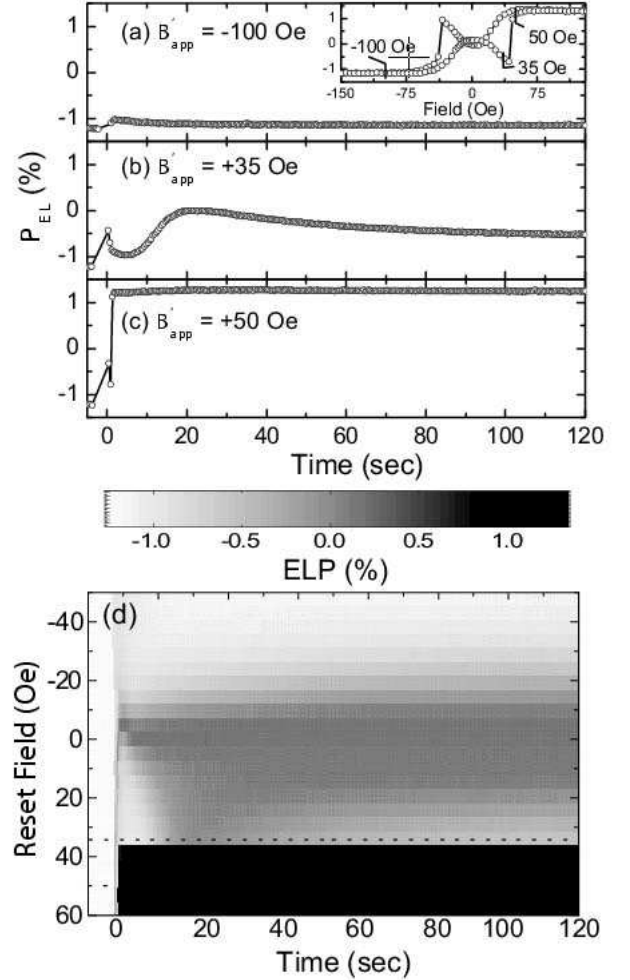}
    \caption[Electroluminescence polarization as a function of time after ramping the field]{\label{fig:CA105LowFieldTimeDep-full}Electroluminescence polarization ($P_{EL}$) as a function of time after rapidly resetting the applied magnetic field from $B_{app}=-500$~Oe to {\bprime} in the oblique easy axis geometry with $\theta=20^\circ$. The field is reset to {\bprime} at $t=0$ at a ramp rate of 1~T/min. Negative times correspond to {\bapp$=-500$~Oe}.   (a)-(c) Single curves of $P_{EL}$ as a function of time after the field is ramped to the reset field {\bprime} as indicated in the figure and by the marks on the oblique easy axis field sweep in the inset of (a). (d) Grayscale intensity map of $P_{EL}$ as a function of reset field {\bprime} and time after reaching {\bprime}.  The curves in (b) and (c) are cross-sections from (d) marked with dotted lines.  The long time-scale response in (b) and at the center of (d) is due to depolarization of the nuclear spin pumped-up at {\bapp}, and simultaneous polarization of nuclear spin at the new field {\bprime}.  The fast response in (c) and at the bottom of (d) represents an adiabatic reversal of nuclear spin as the Fe magnetization switches at the coercive field.}
\end{figure}

A second experimental approach to observing the laboratory scale time dependence of nuclear polarization and depolarization is to rapidly reset {\bapp} while maintaining a constant injected electron spin {\vso}.  The result of such a change in the applied field is a simultaneous polarization and depolarization of nuclear spin in what we will refer to as a \textit{two-spin system} model.  The first spin system is defined as the equilibrium nuclear polarization at {\bapp} prior to resetting the field: $\mIinit(\mbapp)$.  The second spin system is the equilibrium nuclear polarization at the reset field {\bprime}: $\mIprime(\mbprime)$.  At the moment the field is reset, {\Iinit} begins to decay exponentially at {\tonerate}, while {\Iprime} begins to build up according to Eq.~\ref{eq:Iav-BuildUp}.  The average nuclear spin seen by the electrons is the sum of these two spin polarizations:
\begin{equation}\label{eq:Iav-fieldRamping}
    {\bm I}_{av}^\text{\,total}(t)={\bm I}_{av}^\text{\,initial}(t)+{\bm I}_{av}^\prime(t).
\end{equation}
Field reset experiments are most interesting at low applied field, in which either {\bapp} or {\bprime} is less than {\xibl} and the other field is outside the depolarization region.  This requires using the oblique easy axis geometry since in the hard axis configuration {\vso} is a function of {\bapp} at low fields.  

  For this measurement, the magnetic field is initially set well outside the depolarization region at $-500$~Oe, with the sample aligned in the oblique easy axis configuration at $\theta=20^\circ$.  The nuclear spin system is allowed to reach its equilibrium polarization, and then the field is rapidly swept to a new field $|B_{app}^\prime|<200$~Oe at a rate of 167~Oe/second (1~T/minute).  The magnetization is constant over this field range, except for when $B_{app}^\prime$ sweeps past the coercive field for the Fe contact, causing {\mhat} to rotate $180^\circ$.  Time $t=0$ is defined as the point at which $B_{app}^\prime$ is reached.  

Figures~\ref{fig:CA105LowFieldTimeDep-full}(a)--(c) show three traces of $P_{EL}$ versus time after sweeping the field from $B_{app}=-500$~Oe to $\mbprime=-100$, $+35$, and $+50$~Oe. The points on the oblique easy axis hysteresis loop corresponding to these reset fields are marked in the inset to Fig.~\ref{fig:CA105LowFieldTimeDep-full}(a).  The response at $\mbprime=-100$~Oe is essentially flat since the magnetization does not change between $-500$~Oe and $-100$~Oe, and $-100$~Oe is still well beyond the depolarization field {\xibl}.  At $\mbprime=+35\text{~Oe}\sim\mxibl$ the $P_{EL}$ response is long lived and non-monotonic, requiring more than 2 minutes to equilibrate at the EL polarization that corresponds to that of the field swept hysteresis loop in the inset.  The magnetization at $+35$~Oe is the same as at $-100$~Oe, and the $P_{EL}$ response is caused solely by simultaneous nuclear spin depolarization and polarization in response to the reset field being within the depolarization regime near {\xibl}.  At $\mbprime=+50$~Oe, just beyond the coercive field of the Fe contact, $P_{EL}$ discontinuously switches and rapidly equilibrates.  The response at $-45<\mbprime<+60$~Oe is summarized in Fig.~\ref{fig:CA105LowFieldTimeDep-full}(d) as a grayscale intensity map with {\bprime} along the vertical axis and time along the horizontal axis. The $P_{EL}$ scale bar is at the top of the figure.  The dotted lines mark the +35~Oe trace shown in Fig.~\ref{fig:CA105LowFieldTimeDep-full}(b) and the +50~Oe trace in Fig.~\ref{fig:CA105LowFieldTimeDep-full}(c).  The gradual transition from negligible to significant $P_{EL}$ response as {\bprime} moves into the depolarization region, corresponding to the difference between Figs.~\ref{fig:CA105LowFieldTimeDep-full}(a) and (b), is seen in the gray region at the center of Fig.~\ref{fig:CA105LowFieldTimeDep-full}(d).  
The rapid step in $P_{EL}$ when the magnetization switches [Fig.~\ref{fig:CA105LowFieldTimeDep-full}(c)] appears as the nearly constant band of black at the bottom of Fig.~\ref{fig:CA105LowFieldTimeDep-full}(d).  Repeated measurements using a fine scale for {\bprime} through this field range verify that the switching is a discrete event.

The qualitative response of $P_{EL}$ in Fig.~\ref{fig:CA105LowFieldTimeDep-full} can be understood in terms of the two-spin system model of Eq.~\ref{eq:Iav-fieldRamping}.  Before the field is reset, the nuclear spin system is in equilibrium at $\mbapp=-500$~Oe.  When the field is reset to {\bprime}, the initial nuclear spin ${\bm I}_{av}^\text{initial}(\mbapp)$ begins to decay exponentially at a rate $T_1^{-1}$, while nuclear spin ${\bm I}_{av}^\prime(\mbprime)$ builds up according to Eq.~\ref{eq:Iav-BuildUp}.  The average nuclear spin seen by the electrons is the sum of these two components as in Eq.~\ref{eq:Iav-fieldRamping}. Because of the difference between {\tone} and {\tpol}, the sum of the saturating and decaying effective magnetic fields is non-monotonic.  In addition, for {\vbprime} greater than 0~Oe, but less than the coercive field for the Fe contact, the nuclear spin is in a non-equilibrium, negative spin temperature state.  The direction of the nuclear spin polarization is determined solely by {\vs} in Eq.~\ref{eq:I(B)}.  Hence, even when the magnetic field sweeps through zero, the nuclear polarization does not change direction, and at $0<\mbprime<B_\text{coer.}$, $\bm{I}_{av}$ is antiparallel to {\vbprime}.  When {\bprime} moves past the coercive field, the new equilibrium state for nuclear spin in the QW corresponds to a positive spin temperature.  The injected spin exerts a torque on the nuclear spin system and induces a rotation into the equilibrium state while maintaining the magnitude of total nuclear polarization.\cite{Slichter}  This rotation appears as the discrete switching with rapid equilibration in Figs.~\ref{fig:CA105LowFieldTimeDep-full}(c) and (d).  The rotation of nuclear spin is caused by the torque exerted by the Knight field $B_e$.  The Knight field is the effective magnetic field due to spin-polarized electrons that is felt by the nuclei.  For the system investigated here, we expect a Knight field of order $B_e\approx1$~G when the electron spin polarization is 6\%.\cite{Barrett:NMR-in-GaAs-QW-nu=1-PRL1995}

\subsection{\label{sec:Results-NMR}Nuclear Magnetic Resonance}

Nuclear depolarization can also be detected by applying an alternating magnetic field ${\bm B}_1(t)$ perpendicular to {\vbapp} at the resonant frequency for a particular nuclear species, thereby driving a transition between Zeeman eigenstates.\cite{Slichter, Crowell:Strand-spinLED-DNP-NMR-APL2003}
\begin{figure}\begin{center}
  \includegraphics*{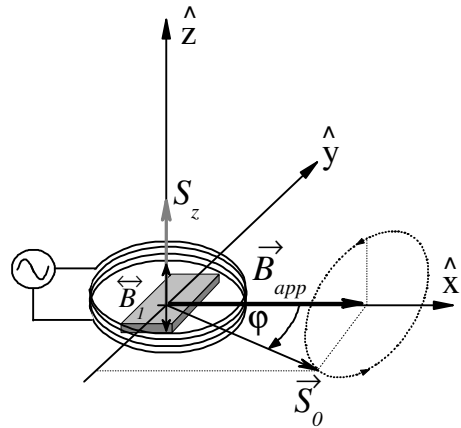}
  \caption[Diagram of nuclear magnetic resonance experimental setup]{\label{fig:NMR-setup}Diagram of the hard axis geometry NMR setup.  {\vbapp} is applied along the magnetic hard axis ($\hat{\bm{x}}$) between 20~Oe and 500~Oe, leading to an angle $\phi$ between {\vso} and {\vbapp}.  A four turn, 1~cm diameter coil, driven by a sinusoidal function generator, produces $\bm{B}_1(t)$ perpendicular to {\vbapp} and parallel to the observation direction.}
\end{center}\end{figure}
The schematic in Fig.~\ref{fig:NMR-setup} shows a diagram of this experiment in the hard axis geometry.  In the presence of a magnetic field, the Zeeman Hamiltonian is
\begin{equation}\label{eq:H-Zeeman}
\mathcal{H}_z=-\gamma\hbar\mvbapp\cdot\bm I,
\end{equation}
where $\gamma$ is the nuclear gyromagnetic ratio.  Taking $\mvbapp=\mbapp\bhat z$, the eigenstates of $\mathcal{H}_z$ have energies $E_z=-\gamma\hbar\mbapp m$ where $m=-\frac{3}{2},-\frac{1}{2},+\frac{1}{2},+\frac{3}{2}$ and the sub-level spacing is $\hbar\omega_0=\hbar\gamma\mbapp$. A magnetic field perpendicular to {\vbapp} oscillating at frequency $\omega_0$ will induce $\Delta m=\pm1$ transitions between the nuclear sub-levels, leading to nuclear spin depolarization.  Because the nuclear depolarization is driven by the coil-generated $B_1$, this will be referred to as \textit{coil-driven NMR}.\cite{Crowell:Strand-spinLED-DNP-NMR-APL2003}

\begin{figure}\begin{center}
    \includegraphics*{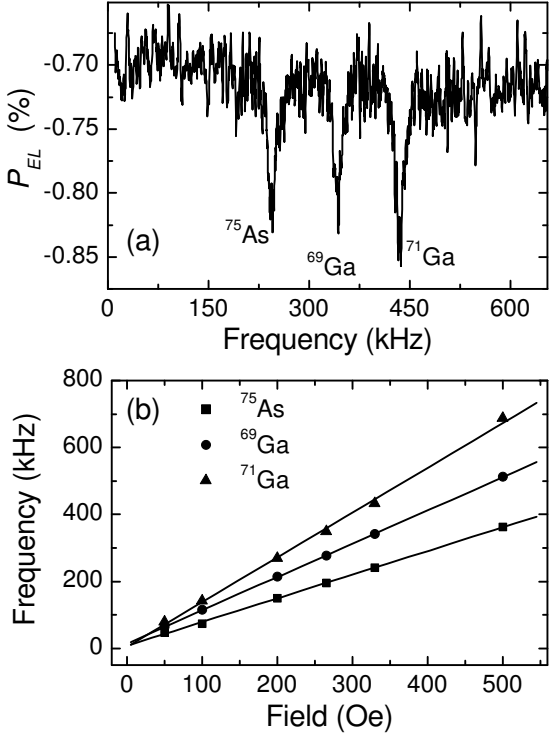}
    \caption[Electroluminescence polarization nuclear magnetic resonance]{\label{fig:S284NMR-simple}(a) Electroluminescence polarization in the hard axis NMR geometry as a function of {\bone} oscillation frequency at $\mbapp=330$~Oe and $\text{T}=20$~K. The depolarization features correspond to resonant nuclear spin depolarization and a reduction in {\Bn}. (b) The frequency of each depolarization feature as a function of {\bapp}.  The slopes of the linear fits give the measured values for the nuclear gyromagnetic ratio $\gamma^{\alpha}$. From Ref.~\onlinecite{Crowell:Strand-spinLED-DNP-NMR-APL2003}.}
\end{center}\end{figure}

Figure~\ref{fig:S284NMR-simple}(a) is a plot of $P_{EL}$ as a function of increasing {\bone} oscillation frequency at fixed {\bapp} in a coil-driven NMR experiment using the setup of Fig.~\ref{fig:NMR-setup}.\cite{Crowell:Strand-spinLED-DNP-NMR-APL2003} The sample is aligned in the hard axis geometry and {\vbone} is along the sample growth direction $\bhat{z}$.  The depolarization signatures for each of the three nuclear isotopes in the GaAs QW: $^{75}$As, $^{69}$Ga, and $^{71}$Ga are easily identified in Fig.~\ref{fig:S284NMR-simple}(a) as sharp dips in $P_{EL}$.  Note that for the data shown in Fig.~\ref{fig:S284NMR-simple}(a) the nuclear spin depolarization results in an \textit{increase} in $P_{EL}$ magnitude.  This is due to the non-monotonic dependence of $P_{EL}$ on {\Bn}, and therefore $I_{av}$, in the hard axis geometry as in Eq.~\ref{eq:S-HardAxis-Nuc}, and reflects a \textit{decrease} in the average nuclear spin polarization.
Figure~\ref{fig:S284NMR-simple}(b) shows the frequency of the peak depolarization for each isotope as a function of {\bapp}.  The lines are linear fits to the data whose slope gives an experimental determination of the gyromagnetic ratio $\gamma^{\alpha}$ for each isotope $\alpha$.  Table~\ref{tbl:S284NMR-gamma} summarizes the measured $\gamma^{\alpha}$, and the accepted values, which agree within experimental error.\cite{CRC}
\begin{table}
\begin{ruledtabular}
\begin{tabular}{rcc}
$\alpha$    &   \multicolumn{1}{c}{measured~$\gamma^\alpha$~(kHz/Oe)}  &   \multicolumn{1}{c}{$\gamma^\alpha$~(kHz/Oe)\cite{CRC}} \\ \hline 
$^{75}$As   &   $0.71\pm 0.10$                 &   0.731 \\
$^{69}$Ga   &   $1.00\pm 0.04$                &   1.025 \\
$^{71}$Ga   &   $1.27\pm 0.04$                &   1.302 
\end{tabular}\end{ruledtabular}
\caption[Nuclear gyromagnetic ratios for GaAs isotopes]{\label{tbl:S284NMR-gamma}Measured and accepted values of the gyromagnetic ratio for all three isotopes present in the GaAs QW.}
\end{table}

\begin{figure}\begin{center}
    \includegraphics*{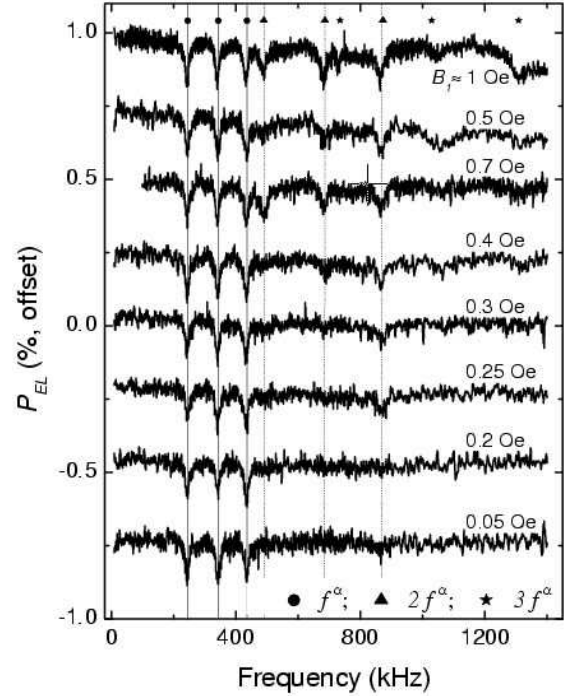}
    \caption[Electroluminescence polarization nuclear magnetic resonance as  a function of {\bone} amplitude]{\label{fig:S284NMR-nonlinear-waterfall}Electroluminescence polarization in the hard axis NMR geometry at $\text{T}=20$~K and $B_{app}=330$~Oe.  Each curve represents a full frequency sweep at a particular peak amplitude for {\bone}, as indicated in the figure.  The symbols and vertical lines mark the positions of the principal NMR resonances, as well as harmonics of the principal resonances. The appearance and amplitude of the dipole-forbidden harmonic resonances are associated with increases in {\bone} amplitude. All curves are offset for clarity.}
\end{center}\end{figure}

The dipole selection rules for the Zeeman eigenstate transitions only allow the $\Delta m=1$ transitions that appear at $f^\alpha=\gamma^\alpha\mbapp$. However, at high $B_1$ amplitude depolarization features are also observed at the higher harmonic frequencies $2f^\alpha$ and $3f^\alpha$.  Figure~\ref{fig:S284NMR-nonlinear-waterfall} shows a sequence of frequency sweeps at increasing $B_1$ amplitude; the data are offset vertically for clarity.  Weak resonance features first appear at $B_{1,\text{peak}}=0.02$~Oe. As the peak amplitude of $B_1$ increases, the primary resonances become stronger, and then the 2$f^\alpha$ and 3$f^\alpha$ transitions appear.  At the amplitude where the 2$f^\alpha$ and 3$f^\alpha$ transitions begin to appear, the primary resonances cease to grow.  While the 2$f^\alpha$ transitions are exactly at twice the primary observed NMR frequencies, the transitions labeled as 3$f^\alpha$ are shifted slightly towards higher frequency.  

\begin{figure}\begin{center}
  \includegraphics*[width=8.5cm]{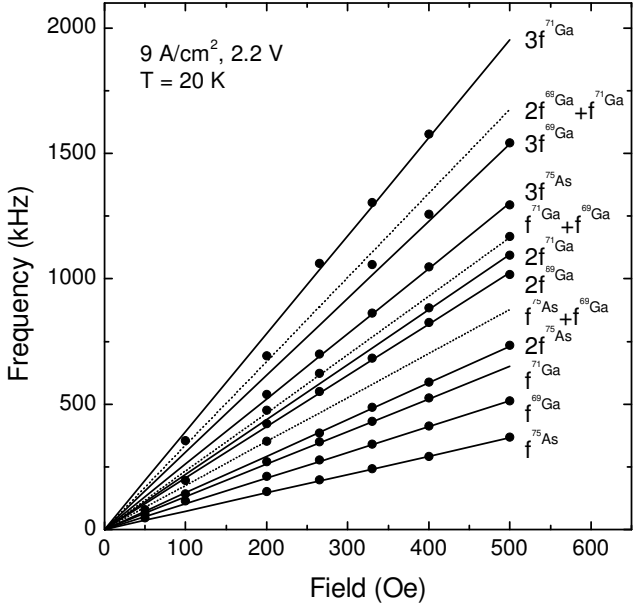}
  \caption[Nuclear magnetic resonance features as a function of applied magnetic field]{\label{fig:S284-NMR-fieldDep}Resonance frequencies as a function of {\bapp} from coil-driven electroluminescence polarization NMR measurements in the hard axis geometry.  The points are the frequencies of resonances from NMR sweeps at fixed field, the lines are calculations based on the allowed Zeeman transitions for the isotopes in the GaAs quantum well, as well as harmonics and sums of those transitions.  The labels along the right hand side refer to the lines.  The field dependence of the additional transitions observed in the electroluminescence polarization NMR identifies these transitions as due to multiples and sums of the principal dipole transitions.  The dotted lines indicate transitions involving coupled nuclei of unlike isotopes.}\end{center}\end{figure}
The field dependence of the additional resonances is shown in Fig.~\ref{fig:S284-NMR-fieldDep}.  The points are from the $P_{EL}$ NMR sweeps, and the lines are calculations of the field dependence for the principal dipole transitions, as well as sums and multiples of those transitions.  The field dependence of the 2$f^\alpha$ and 3$f^\alpha$ transitions clearly establishes their identity.  In addition, there are some resonance features that appear to correspond to sums of unlike isotope transition frequencies, marked with dotted lines.

\begin{figure*}
    \includegraphics*{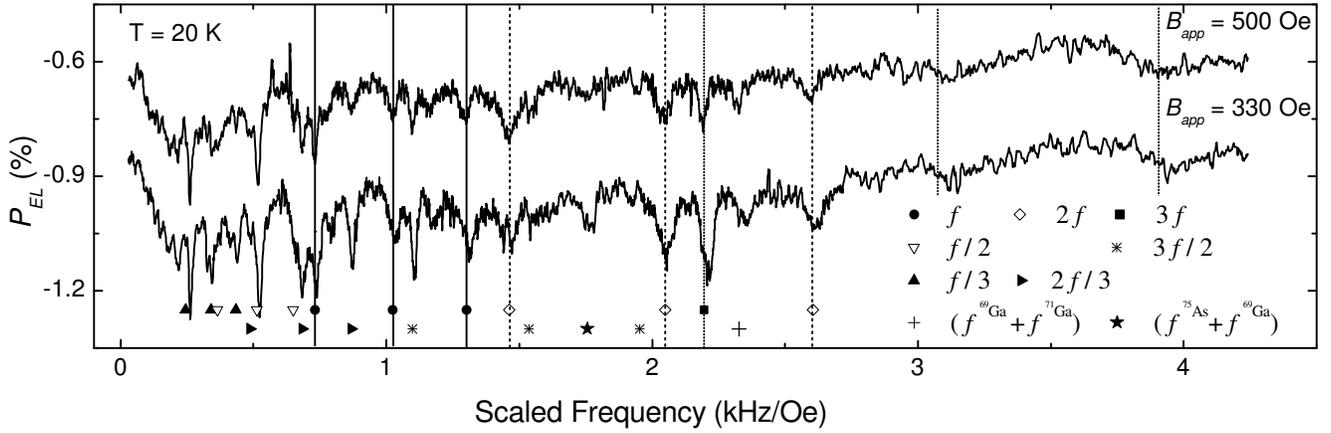}
    \caption[Electroluminescence polarization nuclear magnetic resonance driven by an AC device bias]{\label{fig:S284-NMR-CurrentDriven}Electroluminescence polarization NMR in the hard axis geometry at $\text{T}=20$~K as a function of AC bias frequency; the frequency axis has been scaled in order to compare sweeps from two fields.  The device bias is the sum of a 1.3~V$_\text{DC}$ offset and a 0.65~V$_\text{PP}$ AC sinusoid. The frequency of the AC component is swept at fixed {\bapp} to induce the resonant transitions.  In addition to the principal NMR resonances at $f^\alpha$, resonances are observed at harmonics ($2f^\alpha$ and $3f^\alpha$), sub-harmonics ($\frac{n}{2}f^\alpha$ and $\frac{n}{3}f^\alpha$ where $n=1,2,3$), and at the sum of two different isotope frequencies ($f^{^{75}\text{As}}+f^{^{69}\text{Ga}}$ and $f^{^{69}\text{Ga}}+f^{^{71}\text{Ga}}$). The sub-harmonics are caused by higher harmonic frequencies in the AC bias caused by the diode structure of the device. The sums are unambiguous evidence for strong nuclear spin-spin coupling.}
\end{figure*}

A complementary experiment to the coil-driven NMR of Figs.~\ref{fig:S284NMR-simple}-\ref{fig:S284-NMR-fieldDep} is \textit{current-driven NMR}.  
For the current-driven NMR experiment, the coil of Fig.~\ref{fig:NMR-setup} is removed and the function generator is connected directly to the device.  The function generator output is set for a DC voltage offset below the threshold for EL plus a sinusoidal oscillation. Figure~\ref{fig:S284-NMR-CurrentDriven} shows $P_{EL}$ as a function of the frequency of the AC component of the device bias at $B_{app}=330\text{ Oe, and }500$~Oe; the frequency axis has been scaled by {\bapp} in order to plot both data sets together. The frequencies corresponding to principal and harmonic resonances are marked with vertical lines. There are also prominent transitions between unlike nuclei at frequencies equal to the sum of the two Ga isotopes: $f^{\text{sum1}}=f^{^{69}\text{Ga}}+f^{^{71}\text{Ga}}$ (marked with a $+$), and the sum of $^{75}\text{As}$ and $^{69}\text{Ga}$: $f^{\text{sum2}}=f^{^{75}\text{As}}+f^{^{69}\text{Ga}}$ (marked with a ${\star}$). The sum frequency for $^{71}\text{Ga}$ and $^{75}\text{As}$ is too close to the $2f$ harmonic of $^{69}\text{Ga}$ to clearly resolve the cause of the resonance line ($2f^{^{69}\text{Ga}}\approx f^{^{75}\text{As}}+f^{^{71}\text{Ga}}$), although the depolarization line for this frequency is particularly strong.  

In addition to the sum and harmonic $2f^\alpha$ and $3f^\alpha$ resonances, Fig.~\ref{fig:S284-NMR-CurrentDriven} also exhibits multiple \textit{subharmonic} resonances at $f=\frac{n}{2}f^\alpha\text{, and }\frac{n}{3}f^\alpha$ where $n=1\text{, }2\text{, }3$.  Subharmonics of NMR transitions are expected when there are non-sinusoidal higher harmonics in the AC modulation waveform. This is particularly pronounced in current-driven NMR because the back-to-back diode device structure leads to non-linear current voltage characteristics and therefore to a non-sinusoidal current waveform. 

The depolarization features at sums and multiples of the Zeeman transition frequencies in Figs.~\ref{fig:S284NMR-nonlinear-waterfall}-\ref{fig:S284-NMR-CurrentDriven} demonstrate the presence of perturbations to the Zeeman eigenstates of the nuclei.  There are two primary perturbations that can lead to the higher harmonic transitions: nuclear dipole-dipole coupling and electric quadrupole coupling.  The dipole-dipole Hamiltonian can be constructed from the interaction between two dipoles: $\bm\mu_1=\gamma_1\hbar\bm I_1$ and $\bm\mu_2=\gamma_2\hbar\bm I_2$, and the Hamiltonian can be written as\cite{Slichter}
\begin{equation}\label{eq:H-Dipole}
\mathcal{H}_d=\frac{\gamma_1\gamma_2\hbar^2}{r^3}(A+B+C+D+E+F),
\end{equation}
where
\begin{subequations}\label{eq:H-dipole-A-F}
\begin{eqnarray}
A&=&I_{1z}I_{2z}(1-3\cos^2\theta)\\ \label{eq:H-dipole-A}
B&=&-\frac{1}{4}(I_1^+I_2^-+I_1^-I_2^+)(1-3\cos^2\theta)\\ \label{eq:H-dipole-B}
C&=&-\frac{3}{2}(I_1^+I_{2z}+I_{1z}I_2^+)\sin\theta\cos\theta \text{e}^{-i\phi}\\ \label{eq:H-dipole-C}
D&=&-\frac{3}{2}(I_1^-I_{2z}+I_{1z}I_2^-)\sin\theta\cos\theta \text{e}^{i\phi}\\ \label{eq:H-dipole-D}
E&=&-\frac{3}{4}I_1^+I_2^+\sin^2\theta\text{e}^{-2i\phi}\\ \label{eq:H-dipole-E}
F&=&-\frac{3}{4}I_1^-I_2^-\sin^2\theta\text{e}^{2i\phi} \label{eq:H-dipole-F}
\end{eqnarray}
\end{subequations}
$I_{1,2}^\pm$ are the raising and lowering operators, $\theta$ and  $\phi$ are polar coordinates, and the magnetic field is along $\bhat z$.  The dipole-dipole Hamiltonian enters as a perturbation to the Zeeman Hamiltonian (Eq.~\ref{eq:H-Zeeman}) for two spins,
\begin{equation}\label{eq:H-Zeeman-2spin}
\mathcal{H}_Z=-\gamma_1\hbar\mbapp I_{1z}-\gamma_2\hbar\mbapp I_{2z}.
\end{equation}
The different terms $A-F$ in $\mathcal{H}_d$ connect the nuclear Zeeman sub-levels $m=-\frac{3}{2}\ldots+\frac{3}{2}$.  In the $\mathcal{H}_d=0$ case, the $\Delta m=\pm1$ transitions are allowed while the $\Delta m=\pm2$ and $\pm3$ transitions are forbidden.  However, the terms $B-F$ in the dipole Hamiltonian are off-diagonal and will produce admixtures of the zero-order Zeeman states.  The admixtures have the important effect of making the previously forbidden transitions allowable.  It therefore becomes possible to observe resonances at frequencies two and three times the primary $f^\alpha$, and at sums of frequencies $f^\text{sum}=f^{\alpha1}+f^{\alpha2}$ where $\alpha1$ and $\alpha2$ each correspond to any of the three isotopes in the QW.\cite{Kalevich:twoSpin-NMR-SPSS1982}  

The second perturbation to the Zeeman Hamiltonian that can result in resonances at harmonics of the principal frequencies is the nuclear quadrupole interaction.  The Ga and As nuclei in GaAs are spin-$\frac{3}{2}$ and have electric quadrupole moments. In the presence of an electric field gradient (EFG), the quadrupole moments can affect the allowed transitions for a single nucleus.  The electric field gradients typically associated with quadrupole effects in NMR are not present in GaAs.  However, in {\AlGaAsX}, the partial replacement of Ga atoms with Al atoms alters the local As environment and breaks the cubic symmetry, leading to electric field gradients and quadrupole coupling for the $^{75}$As nuclei.\cite{OO:FleischerMerkulov}  Smaller, but perhaps still significant electric field gradients that affect all three nuclear species will occur at layer interfaces within the heterostructure.   In general, the quadrupole Hamiltonian induces an admixture of Zeeman states that makes $\Delta m=\pm2$ and $\Delta m=\pm3$ transitions allowed for a single nucleus.  However, the EFG for $^{75}$As is larger than EFGs for the other nuclei, and therefore one would expect the $2f^{^{75}\text{As}}$ and the $3f^{^{75}\text{As}}$ signals to dominate those from the Ga isotope harmonics, which is not the case.  Also, significant quadruole coupling should cause splitting in the resonance lines,\cite{Awsch:Salis:OptNMR-PRB2001} which is also not observed.  If quadrupole coupling is a cause of the harmonics in Figs.~\ref{fig:S284NMR-nonlinear-waterfall}-\ref{fig:S284-NMR-CurrentDriven}, then it appears it is a weak coupling that is simply enough to mix the Zeeman eigenstates.

Perturbations to the Zeeman eigenstates from both dipole-dipole coupling and quadrupole coupling can lead to the resonant transitions at $2f^\alpha$ and $3f^\alpha$.  The resonant transitions at $f^\text{sum}=f^{\alpha1}+f^{\alpha2}$ where $\alpha1\neq\alpha2$ in Figs.~\ref{fig:S284-NMR-fieldDep} and \ref{fig:S284-NMR-CurrentDriven} can only be due to $\mathcal H_d$.  Still, there are some inconsistencies between the data and these explanations.  As mentioned above, quadrupole coupling in {\AlGaAsX} should lead to $^{75}\text{As}$ resonances with greater weight than the Ga resonances, which is not observed.  In addition, one would expect the resonance lines to split.\cite{Awsch:Salis:OptNMR-PRB2001}  However, no splittings are observed.  

The second major inconsistency concerns the dipole-dipole coupling. $\mathcal H_d$ is more likely to couple unlike nuclei than like nuclei due to the $1/r^3$ dependence of the dipole interaction, and therefore one would expect more instances of transitions at $f^\text{sum}=f^{\alpha1}+f^{\alpha2}$ where $\alpha1\neq\alpha2$ than are actually observed in coil-driven NMR.  There are very few features in Fig.~\ref{fig:S284-NMR-fieldDep} that correspond to coupling between unlike nuclei and there are none in Fig.~\ref{fig:S284NMR-nonlinear-waterfall}.  In fact, a key distinction between the coil-driven NMR of Figs.~\ref{fig:S284NMR-simple}-\ref{fig:S284-NMR-fieldDep} and the current-driven NMR of Fig.~\ref{fig:S284-NMR-CurrentDriven} is the relative prevalence in the current-driven case of resonances at sum frequencies where $\alpha1\neq\alpha2$.  

A speculative explanation that accounts for the distinction between current-driven and coil-driven NMR with respect to dipole-dipole coupling is that the nuclear spin polarization diffuses away from the QW and into the {\AlGaAs} barriers.  Nuclear spin diffusion has been ignored in all of the preceding discussion of DNP in this paper.\cite{Paget:GaAs-donor-relaxation-PRB1982}  Polarized nuclei in the {\AlGaAs} barriers experience larger electric field gradients due to Al substitution and layer interfaces than the nuclei in the GaAs QW. Therefore, any polarized nuclei in the QW barriers will be comparatively more likely to be perturbed by quadrupole coupling. In the coil-driven NMR experiment, the entire sample is irradiated with the time dependent field {\bone}.  Nuclear spin relaxation in the {\AlGaAs} barriers at frequencies allowed by quadrupole coupling would extend into the QW on the time scale of $T_2\sim100$~$\mu$s. Quadrupole coupling in a coil-driven experiment will allow the $2f^\alpha$ and $3f^\alpha$ resonances that are observed in abundance in Figs.~\ref{fig:S284NMR-nonlinear-waterfall} and \ref{fig:S284-NMR-fieldDep} without the expectation for a similar abundance of $f^\text{sum}=f^{\alpha1}+f^{\alpha2}$ where $\alpha1\neq\alpha2$.  

In the case of current-driven NMR the modulated device bias combines multiple possible drives for nuclear spin transitions, including hyperfine interactions, carrier concentration,\cite{Awsch:Salis:OptNMR-PRB2001, Awsch:Kikkawa:All-opt-NMR-Sci2000} band bending,\cite{Awsch:Poggio:DNP-Gated-NMR-PRL2003} and even Biot-Savart fields from the current. The latter three possible sources would affect the entire sample just as $B_1$ does in coil-driven NMR.  But modulation of the hyperfine interaction is confined to the QW.  The local hyperfine field is equivalent to the Knight shift, which will be approximately 1~Oe in the GaAs QW with $\sim6\%$ electron spin polarization.\cite{Barrett:NMR-in-GaAs-QW-nu=1-PRL1995} A Knight field of a few tenths of an Oe would be enough, in principle, to induce resonant nuclear spin depolarization in the QW, where quadrupole effects should be weakest but dipole-dipole coupling is still present.  The current-driven NMR data of Fig.~\ref{fig:S284-NMR-CurrentDriven} shows  the strongest evidence for dipole-dipole coupled transitions between unlike nuclei, with resonance amplitudes on a similar scale as the other features.  Thus, the difference between coil-driven NMR and current-driven NMR may be the difference between a global drive and a local drive: $B_1$ will interact with all nuclei in the sample while a modulated Knight field will only interact with nuclei in the QW. 

Perhaps the most relevant aspect of the experiment for explaining the resonances at frequencies other than $f^\alpha$ is the amplitude of the depolarizing mechanism.  Note that in both NMR experiments, the appearance of the non-linear features is aided by the strong modulation amplitude of the driving depolarization source: {\bone} in the case of Figs.~\ref{fig:S284NMR-nonlinear-waterfall} and \ref{fig:S284-NMR-fieldDep}, and the device bias in Fig.~\ref{fig:S284-NMR-CurrentDriven}.  Clearly in Fig.~\ref{fig:S284NMR-nonlinear-waterfall} the $2f^\alpha$ and $3f^\alpha$ transitions appear only for high {\bone} amplitudes. These peak values of $B_1$ that generate the non-linear resonance features exceed the typical NMR saturation condition of $(\gamma B_1)^2T_1T_2\sim1$.\cite{Slichter}  In the current case of $\gamma\sim1$~kHz/Oe, $T_1\sim60$~seconds and assuming $T_2\sim100$~$\mu$s, the saturation condition is satisfied for $B_1>0.1$~Oe. All of the points in Fig.~\ref{fig:S284-NMR-fieldDep} are taken at the highest $B_1$ amplitude in Fig.~\ref{fig:S284NMR-nonlinear-waterfall}. Hence the appearance of some non-linear affects may be expected simply from driving the nuclear spin system far out of equilibrium.  Curiously, the primary resonances do not broaden significantly as the saturation condition is exceeded.\cite{Slichter} 

In summary, the NMR data demonstrate saturation of the ordinary magnetic dipole transitions, the existence of dipole-dipole coupling, and that resonant modulation of the spin-polarized current can selectively depolarize the nuclei in the QW.  The presence of a time dependent Knight field in the current-driven case may explain some of the differences observed in the two types of resonance experiments.

\section{\label{sec:Conclusion}Summary and Conclusion}

In this paper we have discussed a series of investigations of electron spin dynamics in the presence of hyperfine effects in Fe/\AlGaAs/GaAs spin injection heterostructures.  The typical nuclear polarization at 20~K in these structures is on the order of 5--10\% and strongly depends on electron spin polarization.  The effective magnetic field felt by the electrons in the configurations studied here is on the order of 4~kG.   

The presence of nuclear polarization is both intrinsic evidence of successful spin-polarized electron transport and provides an additional mechanism for confirming Faraday spin injection measurements.  The electron spin dynamics in the presence of electrically driven nuclear spin polarization are well described by the same set of equations used for dynamics in the presence of optically pumped nuclear spin polarization.  These dynamics can be modeled with four parameters: the steady-state electron spin polarization $S$ in the QW, the conservation of angular momentum in the coupling of the electron and nuclear spin systems as characterized by the leakage factor {\fl}, the precessional field scale {\bhalf}, and $\xi$, the ratio of the relaxation rates for non-hyperfine assisting polarization and depolarization processes.  The upper bound for $S$ is set by Faraday geometry spin injection measurements, {\bhalf} is measured using optical pumping Hanle curves, and single parameter fits to the data on large and small field scales determine {\fl} and $\xi$.  The self-consistency of the bias dependence of $S$ as measured in the Faraday geometry, in which the nuclear polarization does not affect $P_{EL}$, and {\Bn} in the oblique easy axis geometry establishes the accuracy of experimental determinations of electrically injected steady-state spin polarization using the spin-LED.  Finally, the electrically driven NMR and adiabatic inversion of the nuclear spin demonstrate the ability to electrically create and manipulate polarized nuclei using the hyperfine interaction in a ferromagnet/semiconductor heterostructure.

\begin{acknowledgements}
We acknowledge Brian Collins for his early contributions to this experiment.  This work was supported in part by the DARPA SPINS program under ONR/N00014-99-1005 and ONR/N00014-01-0830 and also by the NSF MRSEC program under DMR 98-09364 and DMR 02-12032.
\end{acknowledgements}


\begin{thebibliography}{10}

\bibitem{Lampel:First-opt-DNP}
G. Lampel, Phys. Rev. Lett. {\bf 20},  491  (1968).

\bibitem{OO}
{\em Optical Orientation}, edited by F. Meier and B.~P. Zakharchenya
  (North-Holland Physics Publishers, New York, 1984).

\bibitem{Paget:GaAsNuclearSpinCoupling-PRB1977}
D. Paget, G. Lampel, B. Sapoval, and V.~I. Safarov, Phys. Rev. B {\bf 15},
  5780  (1977).

\bibitem{Ploog:Zhu:first-Fe-spinLED-PRL2001}
H.~J. Zhu, M. Ramsteiner, H. Kostial, M. Wassermeier, H.-P. Schonherr, and
  K.~H. Ploog, Phys. Rev. Lett. {\bf 87},  016601  (2001).

\bibitem{Jonker:Hanbicki:Fe-AlGaAsAPL2002}
A.~T. Hanbicki, B.~T. Jonker, G. Itskos, G. Kioseoglou, and A. Petrou, Appl.
  Phys. Lett. {\bf 80},  1240  (2002).

\bibitem{Jonker:VanTErve:AlO-Schottky2004}
O.~M.~J. {van 't Erve}, G. Kioseoglou, A.~T. Hanbicki, C.~H. Li, B.~T. Jonker,
  R. Mallory, M. Yasar, and A. Petrou, Appl. Phys. Lett. {\bf 84},  4334
  (2004).

\bibitem{Jonker:Li:110-Fe-GaAs-SpinLED-APL2004}
C.~H. Li, G. Kioseoglou, O.~M.~J. {van 't Erve}, A.~T. Hanbicki, B.~T. Jonker,
  R. Mallory, M. Yasar, and A. Petrou, Appl. Phys. Lett. {\bf 85},  1544
  (2004).

\bibitem{IMEC:Motsnyi:OHE-APL2002}
V.~F. Motsnyi, J. {De~Boeck}, J. Das, W. {Van~Roy}, G. Borghs, E. Goovaerts,
  and V.~I. Safarov, Appl. Phys. Lett. {\bf 81},  265  (2002).

\bibitem{IMEC:motsnyi:PRB2003}
V.~F. Motsnyi, P. {Van~Dorpe}, W. {Van~Roy}, E. Goovaerts, V.~I. Safarov, G.
  Borghs, and J. {De~Boeck}, Phys. Rev. B {\bf 68},  245319  (2003).

\bibitem{IMEC:VanDorpe:AlO-Schottky2004}
P. {Van~Dorpe}, W. {Van~Roy}, V.~F. Motsny, G. Borghs, and J. {De~Boeck}, J.
  Vac. Sci. Technol. A {\bf 22},  1862  (2004).

\bibitem{Parkin:Jiang:MTT-spinLED-PRL2003}
X. Jiang, R. Wang, S. {van~Dijken}, R. Shelby, R. Macfarlane, G.~S. Solomon, J.
  Harris, and S.~S.~P. Parkin, Phys. Rev. Lett. {\bf 90},  256603  (2003).

\bibitem{Crowell:Strand-spinLED-DNP-PRL2003}
J. Strand, B.~D. Schultz, A.~F. Isakovic, C.~J. Palmstrom, and P.~A. Crowell,
  Phys. Rev. Lett. {\bf 91},  036602  (2003).

\bibitem{Crowell:Strand-spinLED-DNP-NMR-APL2003}
J. Strand, A.~F. Isakovic, X. Lou, P.~A. Crowell, B.~D. Schultz, and C.~J.
  Palmstrom, Appl. Phys. Lett. {\bf 83},  3335  (2003).

\bibitem{Crowell:Adelmann-spinLED-BiasDep-PRL2004}
C. Adelmann, X. Lou, J. Strand, C.~J. Palmstr{\o}m, and P.~A. Crowell,
  cond-mat/0409103  (2004).

\bibitem{Fiederling:Molenkamp:FirstBeMnZnSe-spinLED}
R. Fiederling, M. Keim, G. Reuscher, W. Ossau, G. Schmidt, A. Waag, and L.~W.
  Molenkamp, Nature {\bf 402},  787  (1999).

\bibitem{OhnoY:Awschalom:FirstGaMaAs_spinLED}
Y. Ohno, D.~K. Young, B. Beschoten, F. Matsukura, H. Ohno, and D.~D. Awschalom,
  Nature {\bf 402},  790  (1999).

\bibitem{Awsch:SpintronicsBook}
{\em Spintronics and Quantum Information Processing}, edited by D.~D.
  Awschalom, D. Loss, and N. Samarth (Springer, Berlin, 2002).

\bibitem{zutic:SpintronicsRMP}
I. Zutic, J. Fabian, and S.~D. Sarma, Rev. Mod. Phys. {\bf 76},  323  (2004).

\bibitem{OO:DyakonovPerel}
M.~I. Dyakonov and V.~I. Perel,  in {\em Optical Orientation}, edited by F.
  Meier and B.~P. Zakharchenya (North-Holland Physics Publishers, New York,
  1984).

\bibitem{en:Value-of-s}
The maximum value of $S$ is $1/2$.  The spin polarization therefore equals $2S$, with a maximum value of 1.

\bibitem{OO:FleischerMerkulov}
V.~G. Fleischer and I.~A. Merkulov,  in {\em Optical Orientation}, edited by F.
  Meier and B.~P. Zakharchenya (North-Holland Physics Publishers, New York,
  1984).

\bibitem{Jonker:ZnMnSe-AlGaAsSpinLED-PRB2000}
B.~T. Jonker, Y.~D. Park, B.~R. Bennett, H.~D. Cheong, G. Kioseoglou, and A.
  Petrou, Phys. Rev. B {\bf 62},  8180  (2000).

\bibitem{SimWindows}
D.~W. Winston, {\em SimWindows Band Structure Simulator, v1.5.0},
  http://www.simwindows.com, 1999.

\bibitem{Sham:Maialle:QW-Exciton-Spin}
M.~Z. Maialle, E.~A. {de~Andrada e Silva}, and L.~J. Sham, Phys. Rev. B {\bf
  47},  15776  (1993).

\bibitem{StonerWohlfarth}
E.~C. Stoner and E.~P. Wohlfarth, Philos. Trans. R. Soc. A {\bf 240},  599
  (1948).

\bibitem{Hanle}
W. Hanle, Z. Phys. {\bf 30},  93  (1924).

\bibitem{Flatte:Lau:spin-relaxation-PRB2001}
W.~H. Lau, J.~T. Olesberg, and M.~E. Flatt\'{e}, Phys. Rev. B {\bf 64},  161301
   (2001).

\bibitem{Flatte:Lau:DP-spin-relaxation-longCondMat2004}
W.~H. Lau, J.~T. Olesberg, and M.~E. Flatt\'{e}, cond-mat/0406201  (2004).

\bibitem{en:eff-g-fctr}
The effective $\mathrm{g}$-factor is estimated to be $\mathrm{g}^*\approx-0.21$ for a {100~\AA} GaAs QW on a [100] substrate based on Ref.~\onlinecite{Malinowsky:Harley-AlGaAs-GaAsQW-gFactor-PRB2000}.  The estimate is provided here as a point of reference since {\bhalf} as measured by the optical pumping Hanle effect already includes the effective {\gfctr}. 

\bibitem{Malinowsky:Harley-AlGaAs-GaAsQW-gFactor-PRB2000}
A. Malinowski and R.~T. Harley, Phys. Rev. B {\bf 62},  2051  (2000).

\bibitem{Overhauser}
A.~W. Overhauser, Phys. Rep. {\bf 92},  411  (1953).

\bibitem{Abragam}
A. Abragam, {\em Principles of Nuclear Magnetism} (Oxford Univ. Press, Oxford,
  1961).

\bibitem{en:Clark-Feher}
For example, due to a large {\gfctr} such as in Ref.~\onlinecite{Clark-Feher:DNP-in-InSb-byDC:PRL1963}.

\bibitem{Clark-Feher:DNP-in-InSb-byDC:PRL1963}
W.~G. Clark and G. Feher, Phys. Rev. Lett. {\bf 10},  134  (1963).

\bibitem{Paget:GaAs-donor-relaxation-PRB1982}
D. Paget, Phys. Rev. B {\bf 25},  4444  (1982).

\bibitem{Flinn:Kerr:1stOptNMR_single_GaAsQW}
G.~P. Flinn, R.~T. Harley, M.~J. Snelling, A.~C. Tropper, and T.~M. Kerr,
  Semiconductor Science and Technology {\bf 5},  533  (1990).

\bibitem{Ohno:Sanada:DNP-in-110-QW-PRB2003}
H. Sanada, S. Matsuzaka, K. Morita, C.~Y. Hu, Y. Ohno, and H. Ohno, Phys. Rev.
  B {\bf 68},  241303  (2003).

\bibitem{Awsch:Poggio:DNP-Gated-NMR-PRL2003}
M. Poggio, G.~M. Steeves, R.~C. Myers, Y. Kato, A.~C. Gossard, and D.~D.
  Awschalom, Phys. Rev. Lett. {\bf 91},  207602  (2003).

\bibitem{Awsch:Salis:OptNMR-PRB2001}
G. Salis, D.~D. Awschalom, Y. Ohno, and H. Ohno, Phys. Rev. B {\bf 64},  195304
   (2001).

\bibitem{Gammon:QD_hyperfine_interaction}
D. Gammon, A.~L. Efros, T.~A. Kennedy, M. Rosen, D.~S. Katzer, D. Park, S.~W.
  Brown, V.~L. Korenev, and I.~A. Merkulov, Phys. Rev. Lett. {\bf 86},  5176
  (2001).

\bibitem{Awsch:Kawakami:FM-imprinting-in-GaAs}
R.~K. Kawakami {\it et~al.}, Science {\bf 294},  131  (2001).

\bibitem{Awsch:Epstein:FPP-PRB2002}
R.~J. Epstein, I. Malajovich, R.~K. Kawakami, Y. Chye, M. Hanson, P.~M.
  Petroff, A.~C. Gossard, and D.~D. Awschalom, Phys. Rev. B {\bf 65},  121202
  (2002).

\bibitem{Awsch:Epstein:FPP-in-nGaAs-PRB2002}
R.~J. Epstein, I. Malajovich, R.~K. Kawakami, Y. Chye, M. Hanson, P.~M.
  Petroff, A.~C. Gossard, and D.~D. Awschalom, Phys. Rev. B {\bf 65},  121202
  (2002).

\bibitem{Awsch:Epstein:FPP-VoltageControl-PRB2003}
R.~J. Epstein, J. Stephens, M. Hanson, Y. Chye, A.~C. Gossard, P.~M. Petroff,
  and D.~D. Awschalom, Phys. Rev. B {\bf 68},  041305  (2003).

\bibitem{Awsch:Epstein:FPP-voltControl-PRB2003}
R.~J. Epstein, J. Stephens, M. Hanson, Y. Chye, A.~C. Gossard, P.~M. Petroff,
  and D.~D. Awschalom, Phys. Rev. B {\bf 68},  041305  (2003).

\bibitem{Awsch:Stephens:Spatial-DNP-by-FPP-PRB2003}
J. Stephens, R.~K. Kawakami, J. Berezovsky, M. Hanson, D.~P. Shepherd, A.~C.
  Gossard, and D.~D. Awschalom, Phys. Rev. B {\bf 68},  041307  (2003).

\bibitem{Awsch:Stephens:Spatial-DNP-Imaging-PRB2003}
J. Stephens, R.~K. Kawakami, J. Berezovsky, M. Hanson, D.~P. Shepherd, A.~C.
  Gossard, and D.~D. Awschalom, Phys. Rev. B {\bf 68},  041307  (2003).

\bibitem{Awsch:Stephens:DNP-doughnuts-APL2004}
J. Stephens, J. Berezovsky, R.~K. Kawakami, A.~C. Gossard, and D.~D. Awschalom,
  Appl. Phys. Lett. {\bf 85},  1184  (2004).

\bibitem{Tarucha:Ono:DNP-by-spinBlockade-QD-PRL2004}
K. Ono and S. Tarucha, Phys. Rev. Lett. {\bf 92},  256803  (2004).

\bibitem{en:Gammon-site}
See Ref.~\onlinecite{Gammon:QD_hyperfine_interaction} for a similar argument with respect to optically driven DNP in quantum dots.

\bibitem{Slichter}
C.~P. Schlicter, {\em Principles of Magnetic Resonance} (Springer, Berlin,
  1991).

\bibitem{Barrett:NMR-in-GaAs-QW-nu=1-PRL1995}
S.~E. Barrett, G. Dabbagh, L.~N. Pfeiffer, K.~W. West, and R. Tycko, Phys. Rev.
  Lett. {\bf 74},  5112  (1995).

\bibitem{CRC}
{\em CRC Handbook of Chemistry and Physics, 76th ed.}, edited by D.~R. Lide
  (CRC Press, Boca Raton, 1995).

\bibitem{Kalevich:twoSpin-NMR-SPSS1982}
V.~K. Kalevich, V.~D. Kul'kov, I.~A. Merkulov, and V.~G. Fleisher, Sov. Phys.
  Solid State {\bf 24},  1195  (1982).

\bibitem{Awsch:Kikkawa:All-opt-NMR-Sci2000}
J.~M. Kikkawa and D.~D. Awschalom, Science {\bf 287},  473  (2000).

\end{thebibliography}
\end{document}